\documentclass[aps,prb,showpacs,twocolumn,showkeys,amssymb,amsmath,superscriptaddress,groupedaddress,reprint]{revtex4-1}

\usepackage[colorlinks=true,linktoc=page,linkcolor=blue,citecolor=red]{hyperref}

\usepackage{amsmath}
\usepackage{braket}
\usepackage{graphicx}
\usepackage{graphics}
\usepackage{color}
\usepackage{stackengine}

\usepackage{verbatim}

\begin{document}

\title{Entanglement contour perspective for strong area law violation in a disordered long-range hopping model}

\author{Nilanjan Roy}
\author{Auditya Sharma}
\affiliation{Department of Physics, Indian Institute of Science Education and Research, Bhopal, Madhya Pradesh 462066, India}


\begin{abstract}
  We numerically investigate the link between the
  delocalization-localization transition and entanglement
   in a disordered long-range hopping model of spinless fermions by studying various static and dynamical
  quantities. This includes the inverse participation ratio,
  level-statistics, entanglement entropy and number fluctuations in the
  subsystem along with quench and wave-packet dynamics. Finite
  systems show delocalized, quasi-localized and localized
  phases. The delocalized phase shows strong area-law violation whereas
  the (quasi)localized phase adheres to (for large subsystems) the strict
  area law. The idea of `entanglement contour' nicely explains the
  violation of area-law and its relationship with `fluctuation contour' reveals a signature at the transition point. The relationship between entanglement entropy
  and number fluctuations in the subsystem also carries signatures for the transition in the
  model. Results from Aubry-Andre-Harper model are compared in this
  context. The propagation of charge and entanglement are contrasted
  by studying quench and wavepacket dynamics at the single-particle
  and many-particle levels.
\end{abstract}

\maketitle

\section{Introduction}
          Ground state wavefunctions of the vast majority of commonly
          encountered Hamiltonians are characterized by the so-called `area-law' of
          entanglement~\cite{hastings2007area,eisert2010,laflorencie2016quantum}. The
          entanglement entropy of a subsystem with respect to its
          complement, scales not as the volume of the subsystem in
          question, but rather as the surface area that links the
          subsystem to its environment. This is loosely justified on
          the grounds that since the couplings are local (for the most
          extensively studied Hamiltonians), quantum correlations in
          the ground state are also local in nature and therefore the
          contributions to the entanglement entropy come from
          correlations at the surface alone. Gapless models show a
          $\log L$-correction to the law~\cite{wolf2006violation,lai2013violation,korepin2004universality}
          - correlations here are stronger than area law because such
          ground states are at a critical point and quantum
          fluctuations induce long-range correlations whereby a region
          deep inside the subsystem offers a non-vanishing
          contribution to correlations with a region far outside
          it. Such mild area-law violations are also fairly
          extensively studied and accepted to be a consequence of the
          criticality of the model. Stronger violations of the
          area-law have also been
          reported~\cite{shiba2014volume,vitagliano2010volume,pouranvari,gori2015explicit}.

	  Long-range couplings are ubiquitous in real physical
          systems, quantum and classical~\cite{campa2014physics,mukamel2009notes}.  A wave
          of current interest exists in uncovering the novel physics
          that emerges when interactions are made
          long-range~\cite{latella2017longrange,mori2013phase,levin2014nonequilibrium}.
          Although the majority of such work is on classical systems,
          there is indeed plenty of interest and work on quantum
          systems. An inexhaustive list includes frustrated
          magnets~\cite{kittel,sandvik2010ground}, spin
          glasses~\cite{young,laumann2014manybody} and various
          ultra-cold atomic~\cite{walker,rajibul,hazzard} and optical
          systems~\cite{sarang}.  One of the characteristics of
          long-range couplings is that, even one-dimensional models
          can give rise to higher-dimensional physics. In quantum
          models, one of the special consequences of this would be
          that by making the couplings to die sufficiently slowly,
          there ought to be stronger violations of the area-law than
          observed in gapless systems. With this hunch in mind, we
          make a detailed study of a long-range disordered hopping
          model in one dimension, where the strength of the couplings
          fall off with distance as a power-law with exponent
          $\sigma$. 
 
                      In the power law model, by tuning the exponent
                      $\sigma$, we are able to discern three distinct phases:
                      one in which the ground state is delocalized and displays
                      a strong area law violation, a second intermediate phase
                      in which the ground state is quasi-localized and adheres
                      to the area law for large subsystem sizes, and a third
                      short range class where the ground state is localized and
                      subscribes to the area-law. The much studied
                      Aubry-Andre-Harper (AAH) model~\cite{aubry,harper} is
                      included for comparison and contrast. The AAH model has
                      the well-known self-dual structure which gives a
                      localization-delocalization transition, with the localized
                      phase being characterized by an area-law abiding
                      entanglement entropy. The quantum phase transition point
                      has the well-known $\log{L}$ correction to the area-law
                      entanglement entropy - we find that in fact, the entire
                      delocalized phase carries the $\log{L}$
                      correction.
         
                   To characterize the phases, we employ several tools
                   including inverse participation ratio (IPR), level
                   spacing ratios, entanglement entropy, subsystem
                   number fluctuations, and non-equilibrium
                   wave-packet dynamics keeping track of the spatial
                   distribution of the wave-packet. For free fermionic
                   models, entanglement entropy has been argued to be
                   closely connected to subsystem number
                   fluctuations\cite{klich2009,song2010general,song2011,HFSong2012,calabrese2012exact,flindt2015}. We
                   find evidence in support of this connection, both
                   in the statics and the dynamics that we study in
                   our model. In this context, we also
                     study a recently introduced quantity called
                     `entanglement contour' which quantifies the
                     contribution from each site in the subsystem to
                     the entanglement. The advantage of this microscopic quantification 
                     is that features like area-law violation and central
                     charge of the system can be obtained from a
                     single subsystem calculation, without the need for any
                     subsystem scaling as with other quantifiers
                     of entanglement~\cite{vidal}. Also its
                     relation with `fluctuation contour' that
                     originates from the number-fluctuations in the
                     subsystem, is useful as a comparative tool~\cite{frerot}. 
                     Entanglement contour nicely captures the the area law and its violation
                     in the disordered long-range hopping model. Also
                     the relationship between the two contours shows
                     striking behavior across the
                     delocalization-localization transition point.
          
                     Non-equilibrium dynamics of a closed quantum system has
                     become a topic of great interest in current
                     research~\cite{Eisert2015,Fagotti,Aditi}. Nowadays one of
                     the key perspectives for understanding different types of
                     phases is the study of entanglement propagation in many-body
                     systems.  This can be probed by tracking quasi-particles in
                     many cases~\cite{zoller2014,cardy2006}.  Also contrasting
                     behavior of various types of transport such as the transport
                     of charge, correlation and entanglement in quantum systems
                     is being used to characterize phases in many-body
                     systems. For example, both the Anderson localized and
                     many-body localized phases show no charge
                     transport~\cite{anderson,baa}; in contrast, the former shows no
                     growth of the bipartite entanglement entropy with time but
                     the latter shows a logarithmic growth~\cite{moore,abanin}. 
                     Recently charge transport and
                     entanglement transport have been contrasted in
                     bond-disordered short-range models~\cite{bera}. We study nonequilibrium dynamics 
                     in our bond-disordered long-range model, finding evidence for the contrast
                     between charge and entanglement propagation.
                     Another aspect of study of long-range models is
                     the generalization of Lieb-Robinson bounds which suggest that in
                     short-range models~\cite{lieb}, the velocity with which correlation spreads is bounded and
                     hence results in a light-cone like spreading of correlation. This leads to a linear
                     growth of entanglement entropy with time following a
                     sudden global quench in short-range models as predicted by
                     related CFT\cite{cardy2006}. The light-cone picture can break down
                     in long-range models; this has been seen theoretically and
                     experimentally in ultracold ion traps for translationally
                     invariant long-range
                     models\cite{Gorshkov,zoller2014,kastner,daley,wouters}. We
                     numerically test the break-down of the light-cone picture
                     in our disordered long-range model and find different
                     results in the delocalized, quasilocalized and localized
                     regimes, which we will discuss later.
                   
                   The paper is organized as follows. In section II, we discuss
                   the delocalization-localization transition in the disordered
                   long-range hopping model.  In section III we explore the
                   entanglement of free fermions in the model at the
                   single-particle and many-particle levels. In subsection IIIA
                   we talk about the single-particle entanglement in the
                   model. In subsection IIIB we study entanglement of fermions
                   and its connection to the number fluctuations in the
                   subsystem. In subsection IIIC we implement the idea of the
                   entanglement and fluctuation contours. In subsection IIID we
                   compare our long-range model with the short-range AAH model. In
                   section IV we investigate the non-equilibrium dynamics
                   at the single-particle and many-particle levels and finally
                   we summarize in section V.

\section{Random long-range hopping model}
We consider a Hamiltonian of the following generic type:
\begin{equation}
\mathcal{H} =  \sum\limits_{i \neq j}^{N} (t_{ij} c_{i}^{\dagger}c_{j} + h.c.) +  \sum\limits_{i}^{N} v_i c_{i}^{\dagger}c_{i},
\label{hamiltonian} 
\end{equation} 
where $c_{i}^{\dagger} (c_{i})$ is the single fermion creation
(annihilation) operator at the $i$th site. In the long-range random
hopping model $t_{ij}=J\frac{u_{ij}}{{r_{ij}}^\sigma}$, is the strength
of hopping and $v_i=0$. $u_{ij}$ is chosen from $[-1,1]$, a
uniform distribution of random numbers and
$r_{ij}=(N/\pi)\sin(\pi|i-j|/N)$, is the geometric chord distance
between the $i^{th}$ and $j^{th}$ sites, when the sites are arranged
in a periodic ring. Here $J$, the maximum magnitude of the hopping
term, is the unit of energy, which we put to unity $J=1$. In a very
similar model \cite{lima1,lima2}, where $r_{ij}={|i-j|}$, $\sigma=1$
has been shown to be the delocalization-localization transition point,
in close connection with the power law random banded matrix (PRBM)
\cite{mirlin1,Cuevas,mirlin2,Evers} model. For $\sigma<1 (\sigma \geq 1)$,
all the eigenstates are delocalized (localized)~\cite{lima1}.
\begin{figure}[h!]
\centering
\includegraphics[width=4.55 cm,height=4.45 cm]{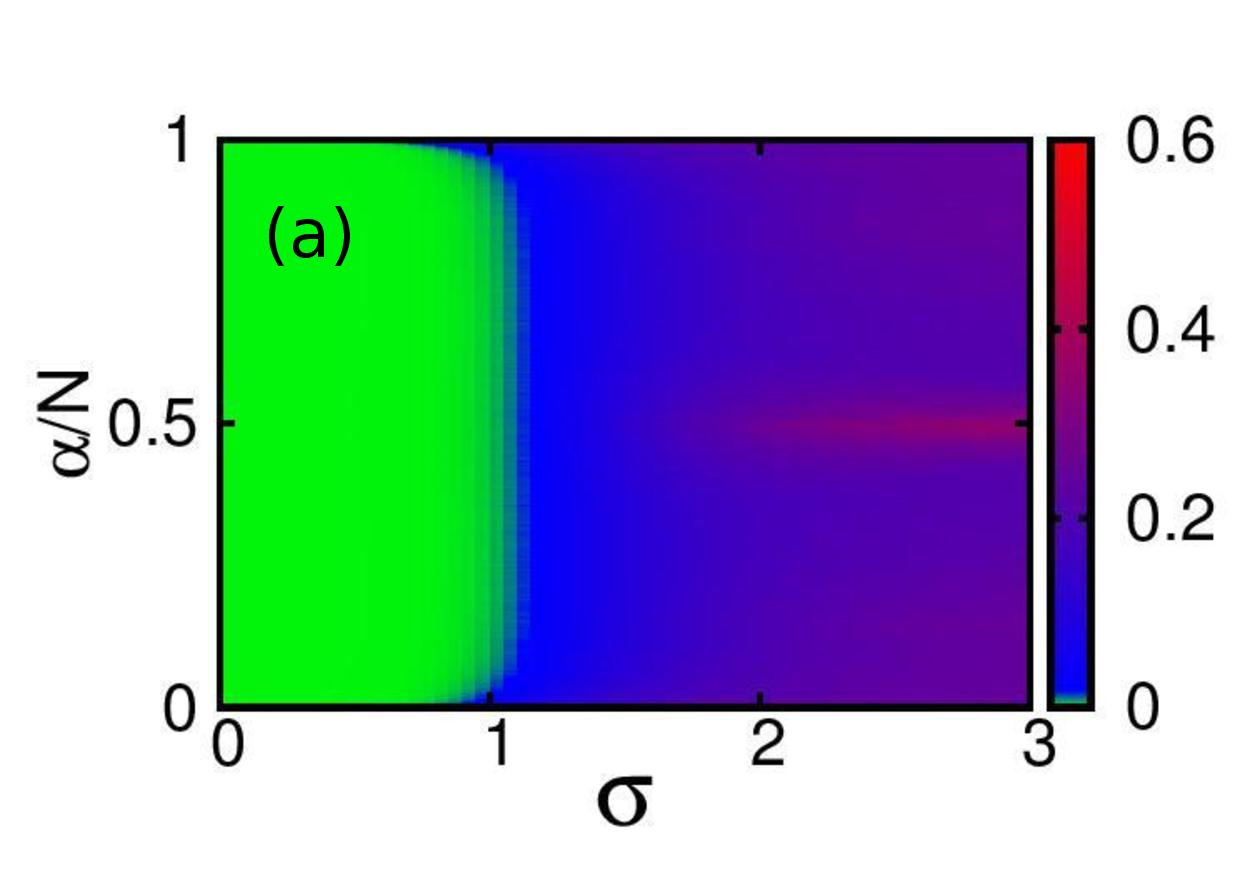}~
\includegraphics[width=4.0 cm,height=3.8 cm]{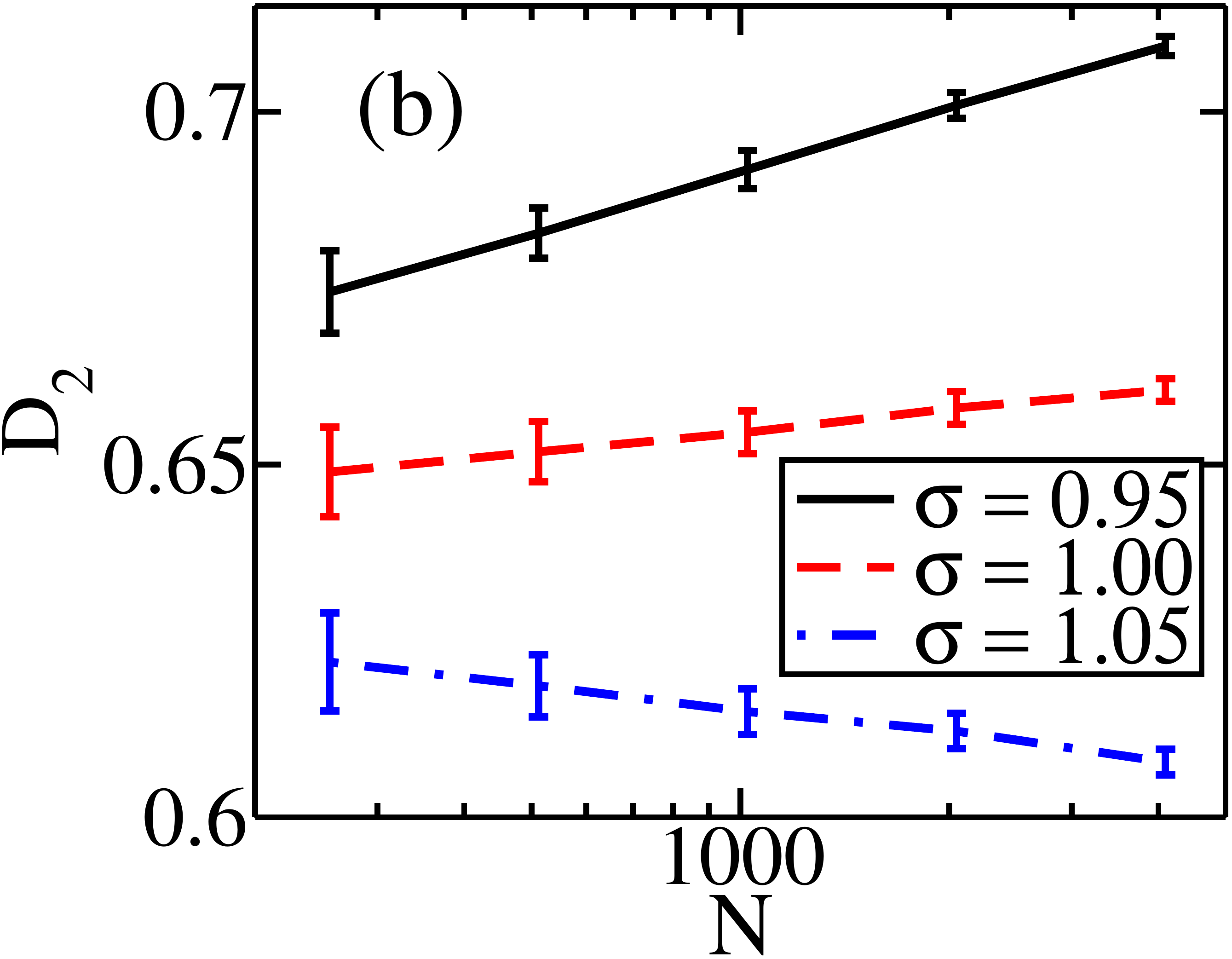}
\includegraphics[width=6.5 cm,height=5.0 cm]{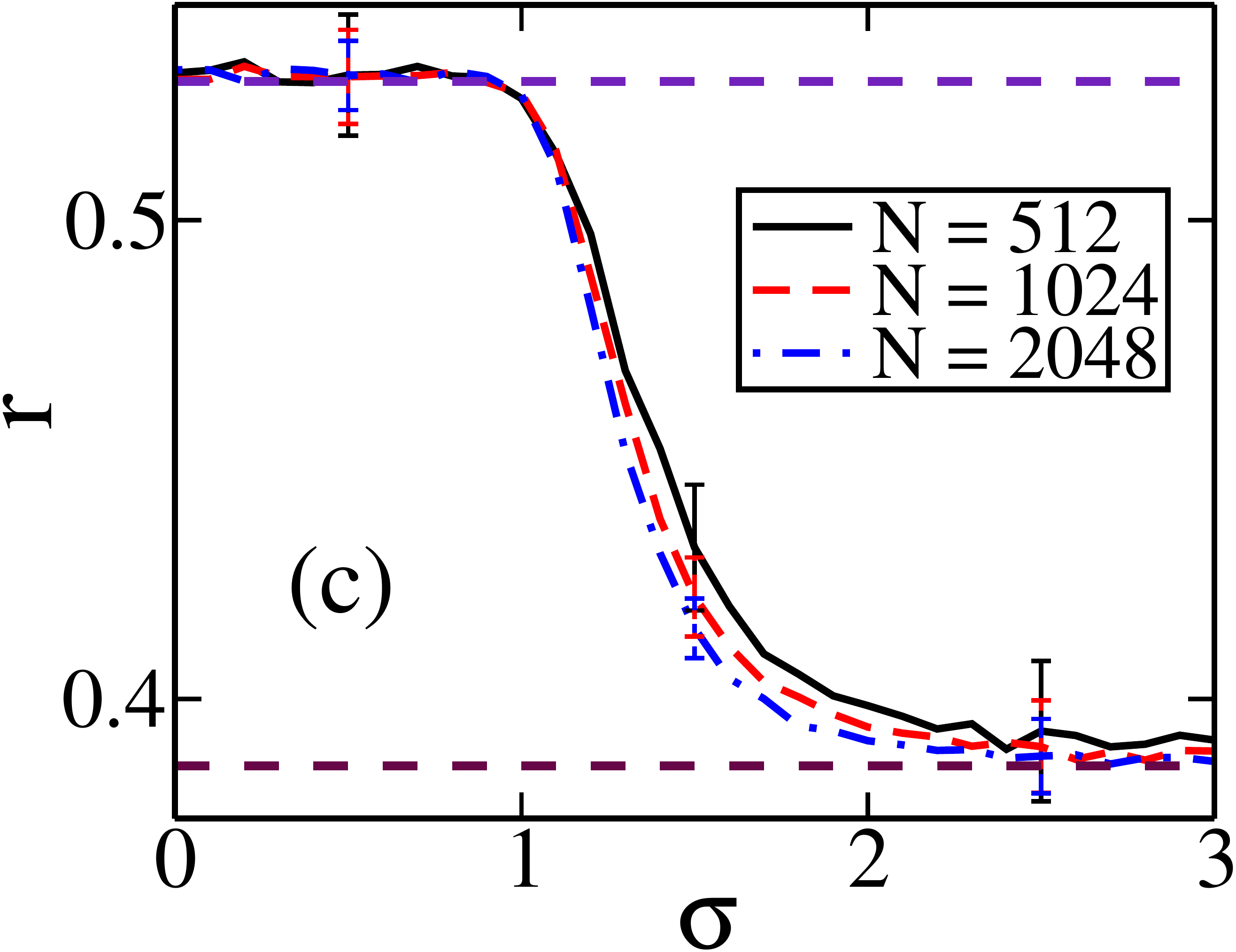}
\caption{(a) Surface plot of IPR of the single particle eigenstates as
  a function of $\sigma$ for system size $N=4096$ and $100$
  realizations of disorder. Here $\alpha$ stands for index of single
  particle eigenstates in ascending order of energy.  The color `green' lies right
  at the bottom of the bar legend, and is barely visible due
  to its very small value ($\sim 1/N$) - it corresponds to 
  the $IPR$ of delocalized eigenstates. (b) Variation of $D_2$ with
  $N$ showing change of slope around $\sigma=1$. (c) The level-spacing
  ratio $r$ as a function of $\sigma$ for increasing system sizes $N$,
  averaged over $100$ realizations of disorder. The two dashed
  horizontal lines denote $r=0.529$ and $r=0.386$ respectively. Error
  bars are of the same order as the samples shown in the three
  regions; for other data points, error-bars are suppressed to enhance
  clarity. }
\label{transition}
\end{figure}

To quantify the point of the localization transition, we compute the inverse participation ratio (IPR), which is defined as 
\begin{equation}
I(\alpha) = \sum\limits_{i=1}^{N} {|\psi_i(\alpha)|}^4,
\end{equation}
where the coefficients are drawn from the $\alpha^{th}$ normalized single particle eigenfunction $\ket{\psi(\alpha)}=\sum\limits_{i}^{} \psi_i(\alpha) \ket{i}$ expanded in the complete set of the Wannier basis $\ket{i}$, which represents the state of a single particle localized at the site $i$ of the lattice. IPR of all the eigenstates as a function of $\sigma$
is shown in the surface plot Fig.~\ref{transition}(a). We see the presence of
localized states at the edges of the band near $\sigma=1$, which is essentially a finite size effect \cite{lima1}.
\begin{figure*}
\centering
\includegraphics[width=5.7 cm,height=4.2 cm]{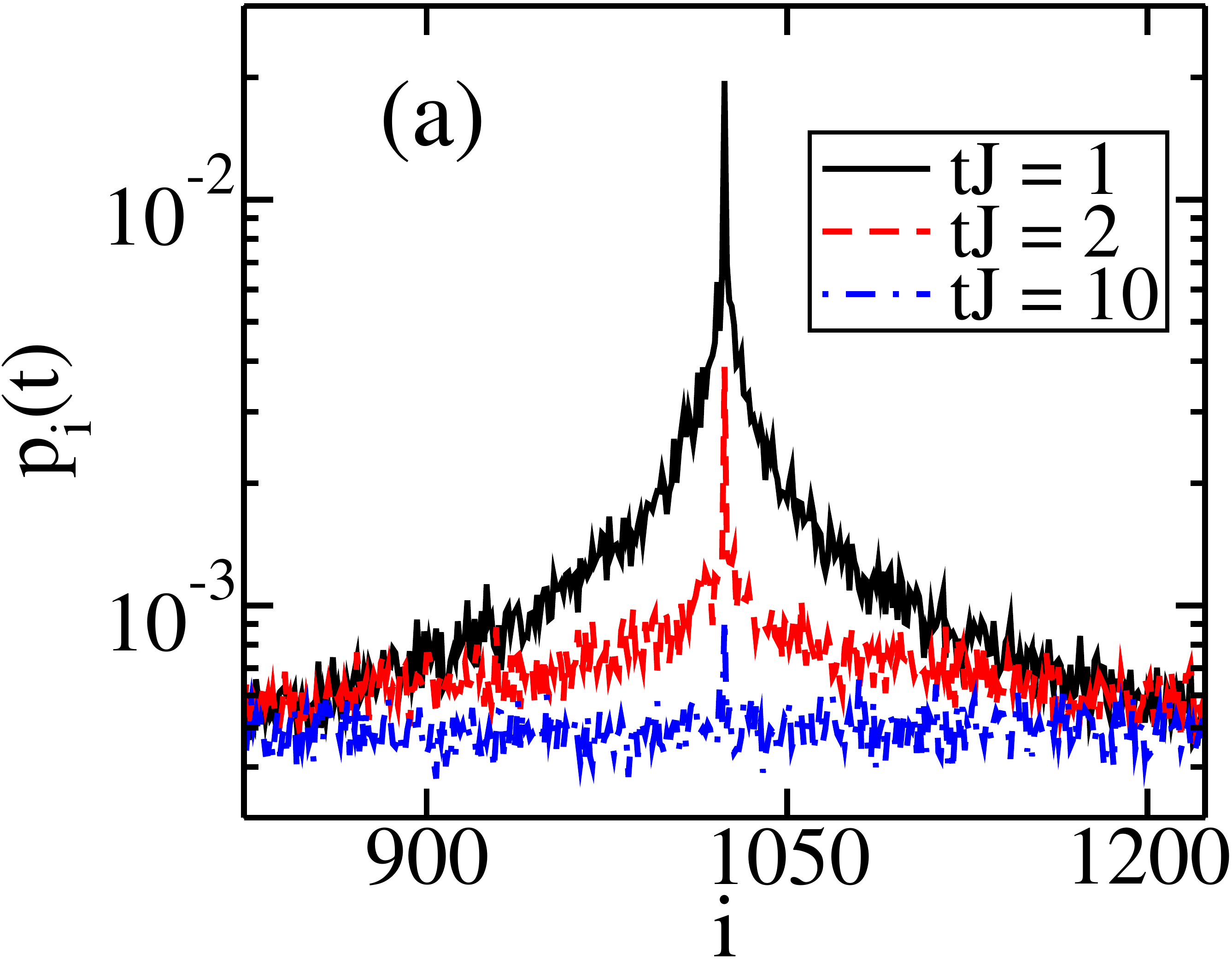}
\includegraphics[width=5.7 cm,height=4.2 cm]{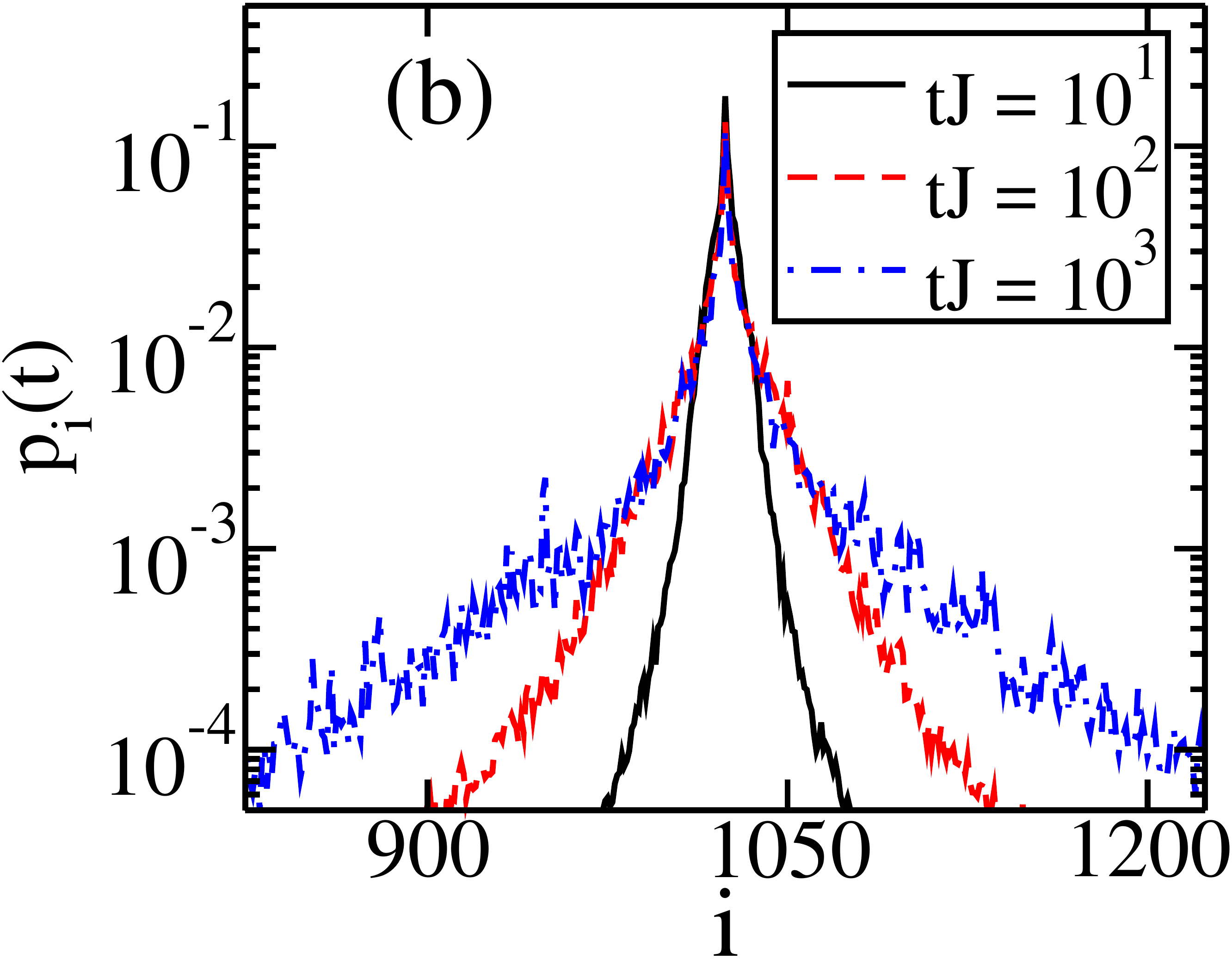}
\includegraphics[width=5.7 cm,height=4.2 cm]{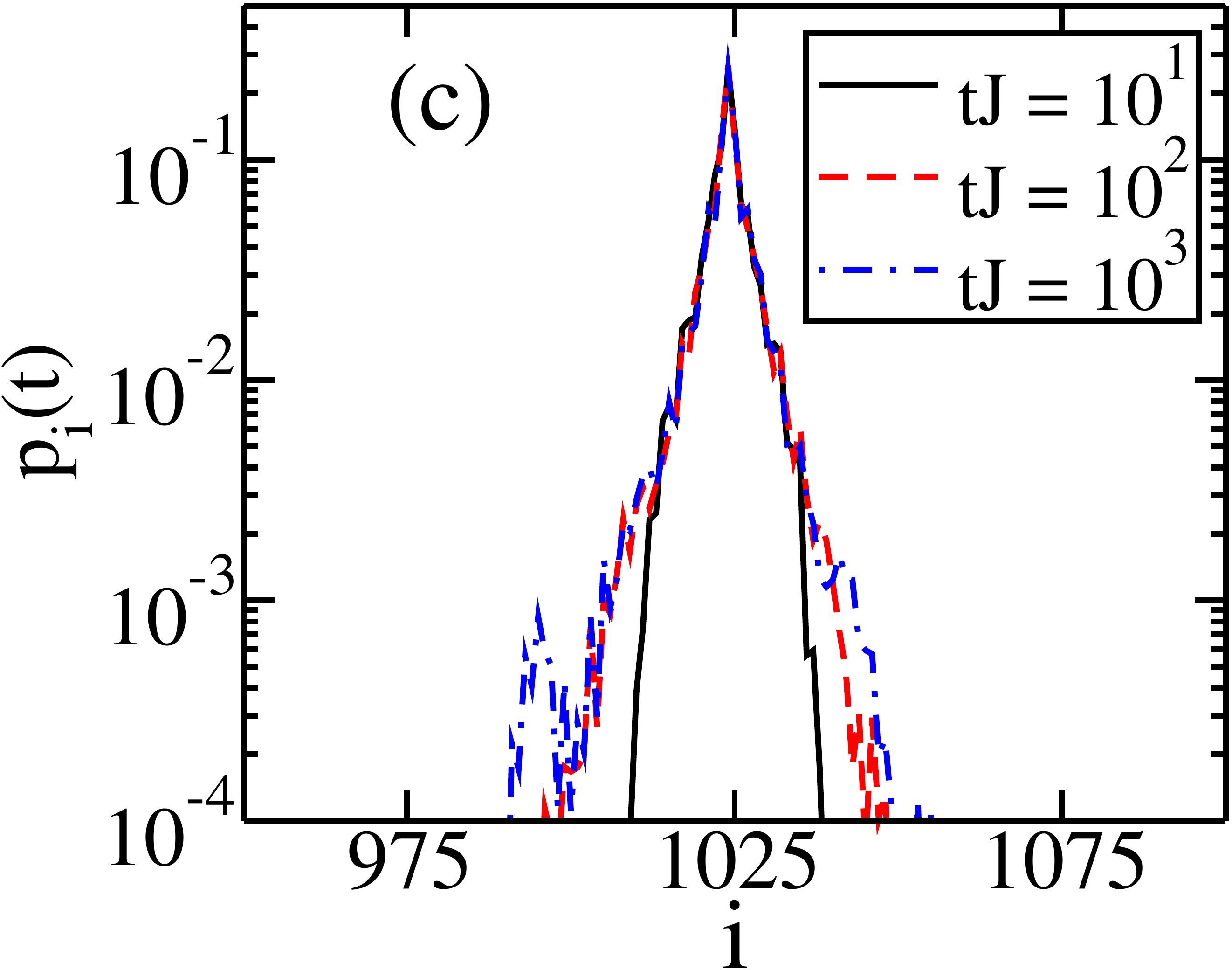}
\caption{ Probability distribution $p_i(t)$ for finding a single particle, initially at the middle of the lattice, at each site of the lattice for increasing values of time (in units of $J^{-1}$) (a) for $\sigma=0.5$ (delocalized phase); (b) for $\sigma=1.5$ (quasilocalized phase); and (c) for $\sigma=3.0$ (localized phase) respectively. For all the plots $N=2048$ and number of disorder realizations is $100$. }
\label{wfn}
\end{figure*}

We also calculate the participation moments averaged over all the
eigenstates. The $q^{th}$ participation moment is obtained by averaging over all the eigenstates and disorder configurations
\begin{equation}
P_q=\bigg \langle \frac{\sum\limits_{\alpha=1}^{N} {P_q}(\alpha)}{N} \bigg \rangle,
\end{equation}
where ${P_q}(\alpha)=1/\sum\limits_{i=1}^{N}|{\psi_i(\alpha)|}^{2q}$. However, $P_q\propto N^{D_q(q-1)}$. In a fully delocalized (localized) regime $D_q$ approaches unity (zero) as the thermodynamic limit is approached. It is evident that $\frac{\log P_q}{\log N} \propto D_q (q-1)$ and from the variation of $D_q$ with the system size, one can identify the point of transition in the thermodynamic limit. We choose $q=2$ and $D_2$ is plotted with the system size in Fig.\ref{transition}(b). The $D_2$ vs $N$ plot changes slope at $\sigma=1$, which is the point of the localization transition.

The mean of the ratio~\cite{huse2007,Atas2013} $r$ between adjacent gaps ($\delta$) in the spectrum can be used to identify a crossover from Wigner-Dyson statistics in the delocalized phase to Poisson statistics in the localized phase. Defining
\begin{equation}
r_k = \frac{\min(\delta_k,\delta_{k+1})}{\max(\delta_k,\delta_{k+1})},
\end{equation}
where $\delta_k = \epsilon_{k+1}-\epsilon_{k}$ is the $k^{th}$ energy gap, the mean
ratio is $r = \langle \overline{r} \rangle$, where the bar represents
an average over the spectrum, and the angular brackets the average
over disorder. It is known from random matrix theory that the mean
ratio $r$ is approximately $0.529$ in the delocalized phase and
$0.386$ in the localized phase~\cite{huse2007,Atas2013}.
Fig.~\ref{transition}(c), based on the finite sizes considered here,
suggests that the system is in the ergodic phase in the region $0 \leq
\sigma \leq 1$.  Then $r$ starts decreasing till it reaches the
localized phase around $\sigma=2$.  The intermediate phase showing
intermediate distributions is discussed in the following analysis.

In order to better understand the presence of different phases in the system, we have considered a wavepacket initially localized at the middle site $i_0$ of the lattice i.e. $\psi_i(t=0)=\delta_{i,i_0}$ and calculated the evolution of the spatial distribution of the wavepacket with time. The probability of finding a particle at site $i$ at a given instant $t$ is given by $p_i(t)=|\psi_i(t)|^2$. The spatial dependence of the probability distribution for increasing time is shown in Fig.\ref{wfn}. It is to be noted that in the quasilocalized phase (Fig.\ref{wfn}(b)), the central part of the wavepacket rapidly drops down to a smaller value, which then barely changes with time whereas the tails of the wavepacket keep spreading with time. In the delocalized phase, the occupancy at the initial site along with all the other sites rapidly decreases and the wavepacket takes the form of a uniform distribution(Fig.\ref{wfn}(a)) whereas in the localized phase, the dynamics of the wavepacket is almost absent and it becomes almost exponentially localized (Fig.\ref{wfn}(c)). Fig.~\ref{wfn} thus shows that the quasi-localized phase is distinct from both the delocalized and localized phases, and yet carries some character of each of these phases.
\section{Entanglement in the model}
Phase transitions in extended quantum systems are
known to be captured by different measures of entanglement~\cite{horodecki2009quantum,amico2008entanglement,eisert2010} such as
concurrence, entanglement entropy etc. In the subsequent part of this
section, we will calculate the von Neumann entanglement entropy
between a suitable subsystem and its complement, both for single particle and many particle
states. We will investigate if there is a violation of the `area law'
of the entanglement entropy and analyze our results on the basis of
the localization transition. We will discuss local
particle-number fluctuations and its relation with entanglement
entropy in the context of the transition in our model. Also  we discuss the `entanglement contour' and `fluctuation contour' in this context.  
\subsection{Single-particle entanglement}
First we discuss single-particle entanglement entropy, which has been argued to be a useful resource for quantum information processing~\cite{vanEnk2005,Dashenbrook2016}. In order to calculate the entanglement entropy between two subsystems A and B for the normalized single particle states, one writes down a normalized single particle eigenstate in the following way
\begin{equation}
\ket{\psi}=\sum\limits_{i\in A} \psi_i {c_i}^\dagger \ket{0}_A\otimes\ket{0}_B + \sum\limits_{i\in B} \psi_i\ket{0}_A\otimes{c_i}^\dagger\ket{0}_B, 
\end{equation}
where $\ket{0}_{A/B}$ is the vacuum state in the subsystem A/B. Then the reduced density matrix $\rho^{sp}_A=Tr_B(\ket{\psi}\bra{\psi})$ has two eigenvalues $p_A=\sum\limits_{i \in A} |\psi_i|^2$ and $p_B=1 - p_A$\cite{Chakravarty}(see Appendix \ref{appA} for more details). The single particle entanglement entropy is then given by 
\begin{equation}
S^{sp}_A=-p_A\ln p_A - p_B\ln p_B.
\end{equation}
\begin{figure}
\centering
\includegraphics[width=4.275 cm,height=4.0 cm]{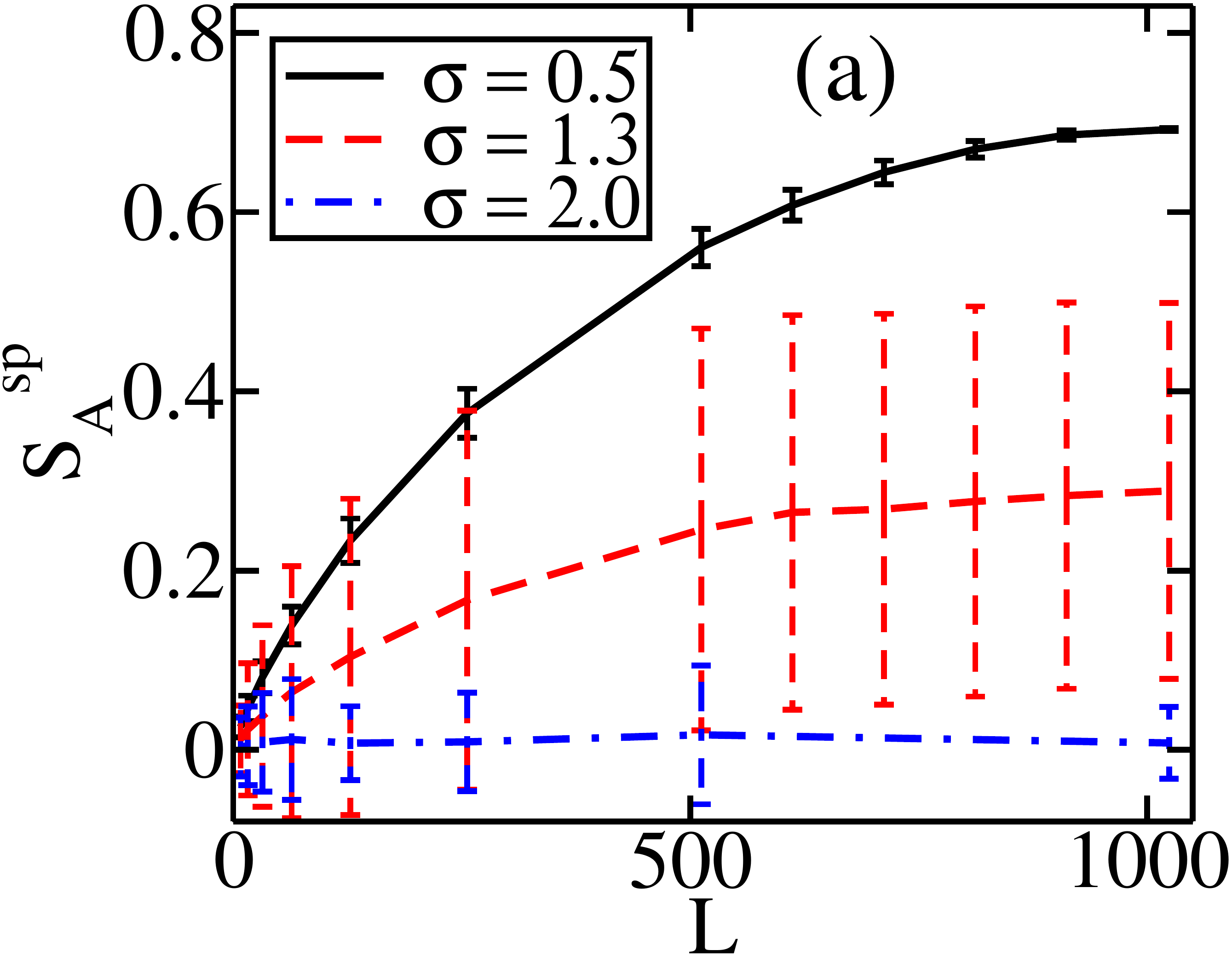}
\includegraphics[width=4.275 cm,height=4.0 cm]{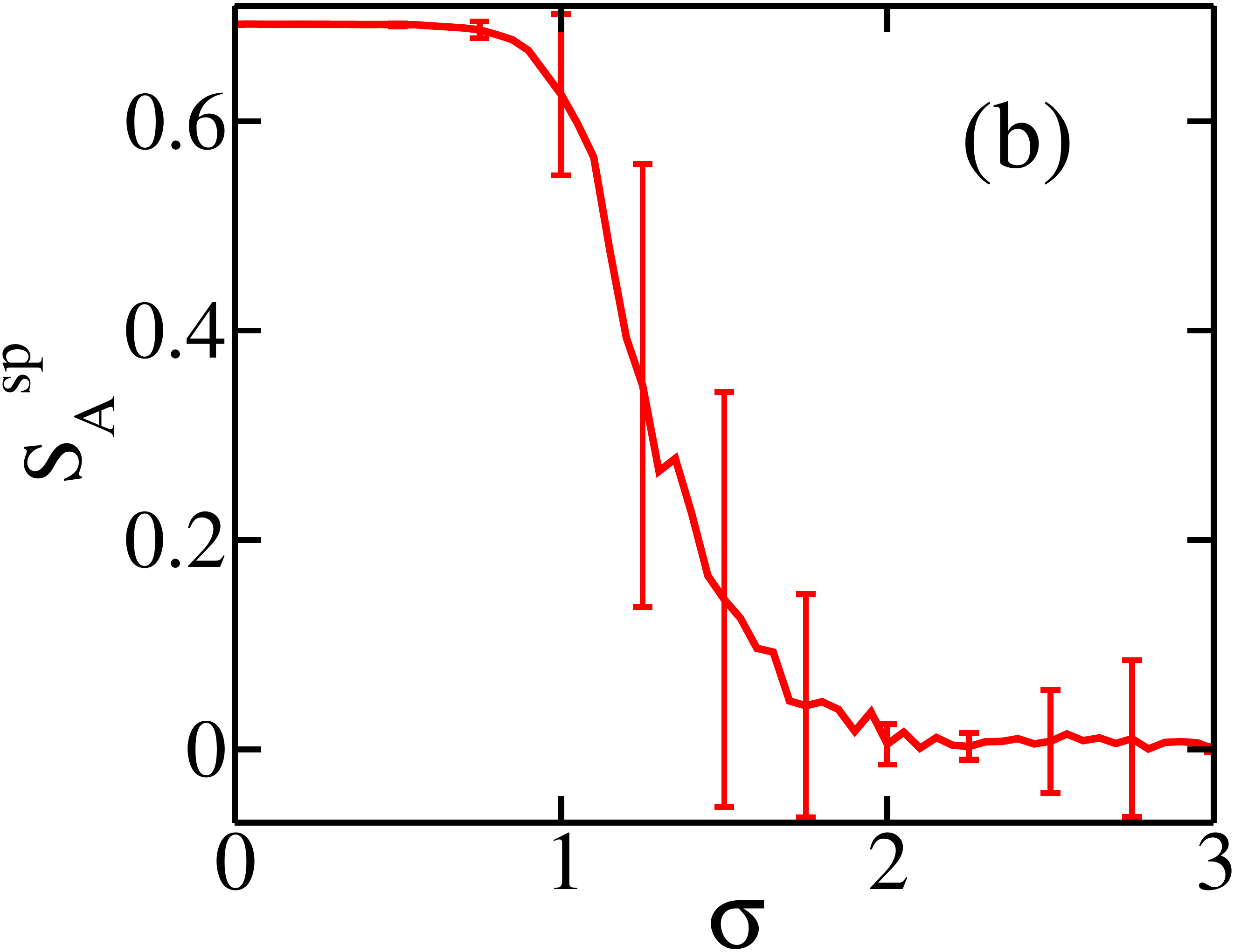}
\caption{(a) Scaling of single-particle entanglement entropy $S^{sp}_A$ with subsystem size $L$  for different values of $\sigma$. (b) Variation of $S^{sp}_A$ with $\sigma$, where $L=N/2$. The system size $N=2048$ and number of disorder realizations is $100$ for both the plots.}
\label{singleEE}
\end{figure} 
This entropy is bounded between $\ln2$ and $0$. In a delocalized
eigenstate, $S^{sp}_A$ increases with $L$, the size of the subsystem
A, as $p_A=L/N$ and reaches the maximum value $\ln2$, when $L=N/2$. In
a single site localized state $S^{sp}_A$ is $0$ as $p_A=1$ or $0$ and
does not show any variation with the subsystem size. The variation of
$S^{sp}_A$ with $L$ in different phases for our model is shown in
Fig.\ref{singleEE}(a). In the quasi-localized phase, $S^{sp}_A$ varies
with $L$ but its maximum value is less than $\ln2$ and the maximum
value decreases as $\sigma$ increases towards $\sigma=2$. The curves
deviate more from the delocalized ones as $L$ increases towards $N/2$
because in the quasilocalized eigenstate the central part of the
wavefunction is more localized compared to the tails. The variation of
$S^{sp}_A$ with $\sigma$ can be seen from Fig.\ref{singleEE}(b). The
delocalized $(\sigma<1)$, quasi-localized $(1<\sigma<2)$ and localized
$(\sigma>2)$ phases are clearly seen from the plot. Also it is worth
mentioning that the quasilocalized phase shows large intrinsic
fluctuations in $S^{sp}_A$. This results in large error-bars that
cannot be significantly reduced by increasing the number of disorder
realizations. This is obvious because in the quasi-localized phase,
for an eigenstate, the probability distribution for finding a single
particle has multiple peaks and they can appear in random places in
the lattice for different realizations of disorder (not shown here)
thus making $p_A$ a highly fluctuating quantity. In the localized
phase the probability distribution is more or less singly peaked hence
$p_A$ is always close to $0$ or $1$ whereas in the delocalized phase
the probability distribution has no peak and it is a uniform one,
hence $p_A\sim L/N$ giving rise to smaller error-bars in $S^{sp}_A$.

\subsection{Fermionic entanglement and fluctuations}   
In this subsection we consider noninteracting spinless fermions at
half-filling in the system and investigate signatures of the
localization transition via entanglement in many-body states. The connection between localization and entanglement is subtle. Intuitively, one would expect that the greater the delocalization, the more the entanglement and vice versa; however, this correlation is not absolute and counterexamples are available\cite{kannawadi2016persistent}. We also discuss the
relationship between subsystem number fluctuations and entanglement entropy in the model. We start with a brief discussion of the calculation of the
entanglement entropy of fermions in the
ground state\cite{peschel2003calculation,peschel2009,peschel2012special}(see
Appendix \ref{appB} for details). For the fermionic many-body ground state
$\Ket{\Psi_0}$, the density matrix can be written as
$\rho=\Ket{\Psi_0} \Bra{\Psi_0}$. The entanglement
entropy between two subsystems is then given by $S_A=-Tr(\rho_A \log
\rho_A)$, where the reduced density matrix
$\rho_{A}=Tr_{B}(\rho)$. However, for a single Slater determinant
ground state, Wick's theorem can be exploited to write the reduced
density matrix as $\rho_{A}=\frac{e^{-H_{A}}}{Z}$, where
$H_{A}=\sum\limits_{ij} H_{ij}^A c_{i}^{\dagger}c_{j}$ is called the
entanglement Hamiltonian, and $Z$ is obtained from the condition $Tr
(\rho_{A}) = 1$. The information contained in the reduced density
matrix of size $2^L\times 2^{L}$ can be captured in terms of the
correlation matrix $C$ of size $L\times
L$\cite{peschel2003calculation} within the subsystem A, where
$C_{ij}=\left\langle c_{i}^{\dagger}c_{j} \right\rangle $. The
correlation matrix and the entanglement Hamiltonian are related
by\cite{peschel2003calculation,peschel2009,peschel2012special}:
\begin{equation}
  \label{eqn:C_and_H}
  C=\frac{1}{e^{H_A}+1}.
\end{equation} 
Using this relation, the entanglement entropy for free fermions is given by\cite{peschel2009,peschel2012special},
\begin{equation}
S_A=-\sum\limits_{m=1}^{L} [\lambda_m \log \lambda_m + (1-\lambda_m) \log (1-\lambda_m)],
\end{equation}
where $\lambda_m$'s are the eigenvalues of the correlation matrix $C$. 
\begin{figure}
\includegraphics[width=6.5 cm,height=5.5 cm]{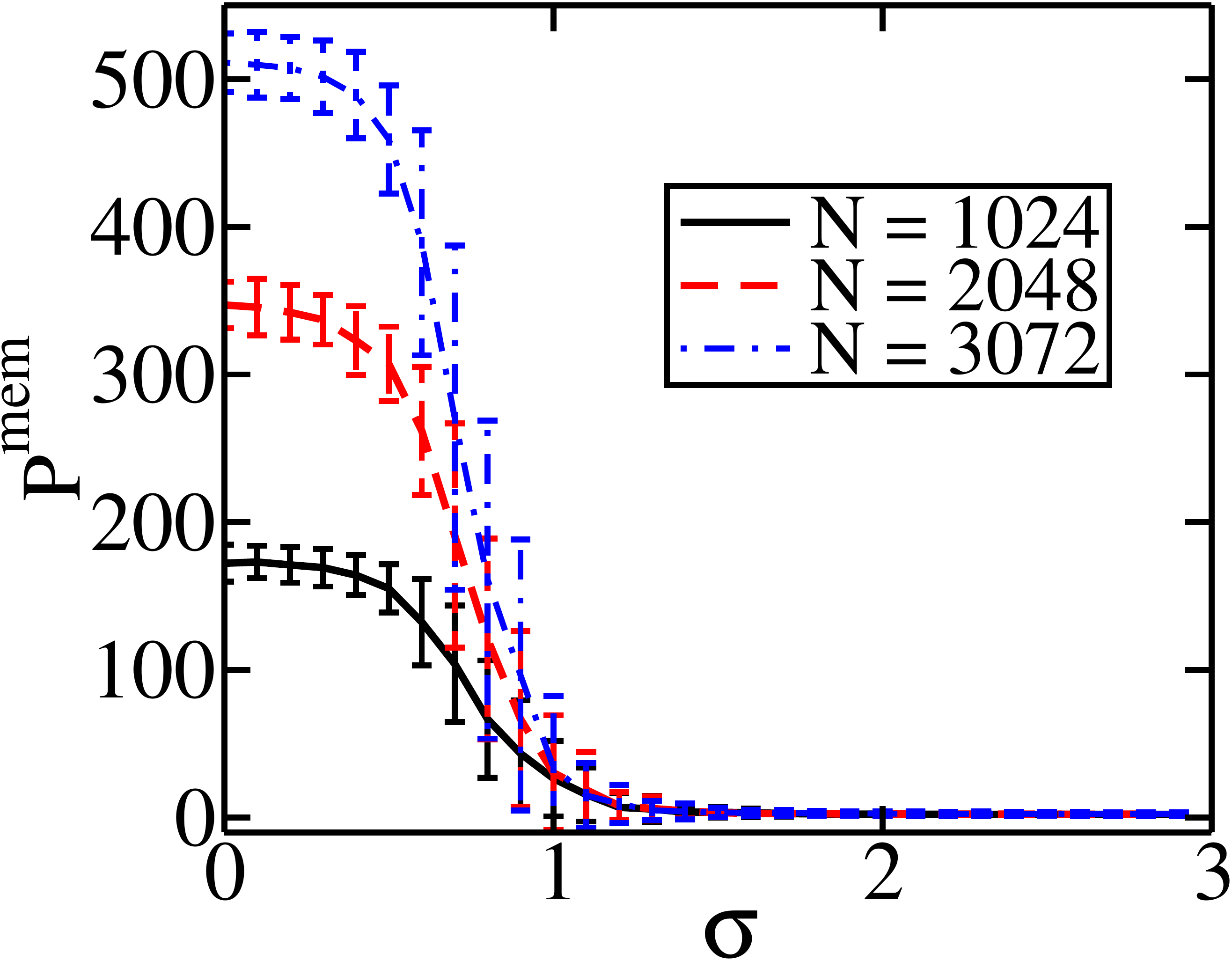}
\caption{Participation ratio of MEM, denoted as $P^{mem}$, as a function of $\sigma$ for increasing system sizes $N$, averaged over $100$ disorder realizations. Subsystem size $L=N/2$ for fermions at half-filling.}
\label{mem}
\end{figure}
It has been conjectured that the zero mode of the entanglement
Hamiltonian has information about topological quantum phase
transitions~\cite{LiHaldane}. The same conjecture can be extended to a
non-topological system \cite{pouranvari}. It follows from
Eqn.~\ref{eqn:C_and_H} that the zero mode of the entanglement
Hamiltonian would correspond to the eigenfunction of the correlation
matrix, whose eigenvalue is equal (closest) to $0.5$ . As this
eigenmode contributes the maximum to the entanglement entropy, it is
called the maximally entangled mode (MEM). The participation ratio of
the MEM reflects the localization transition at $\sigma=1$
[Fig.~\ref{mem}]. This is a nice example of detecting the localization
transition from the entanglement spectra without having any prior
knowledge about the original Hamiltonian.

\begin{figure}
\centering
\includegraphics[width=6.5 cm,height=5.0 cm]{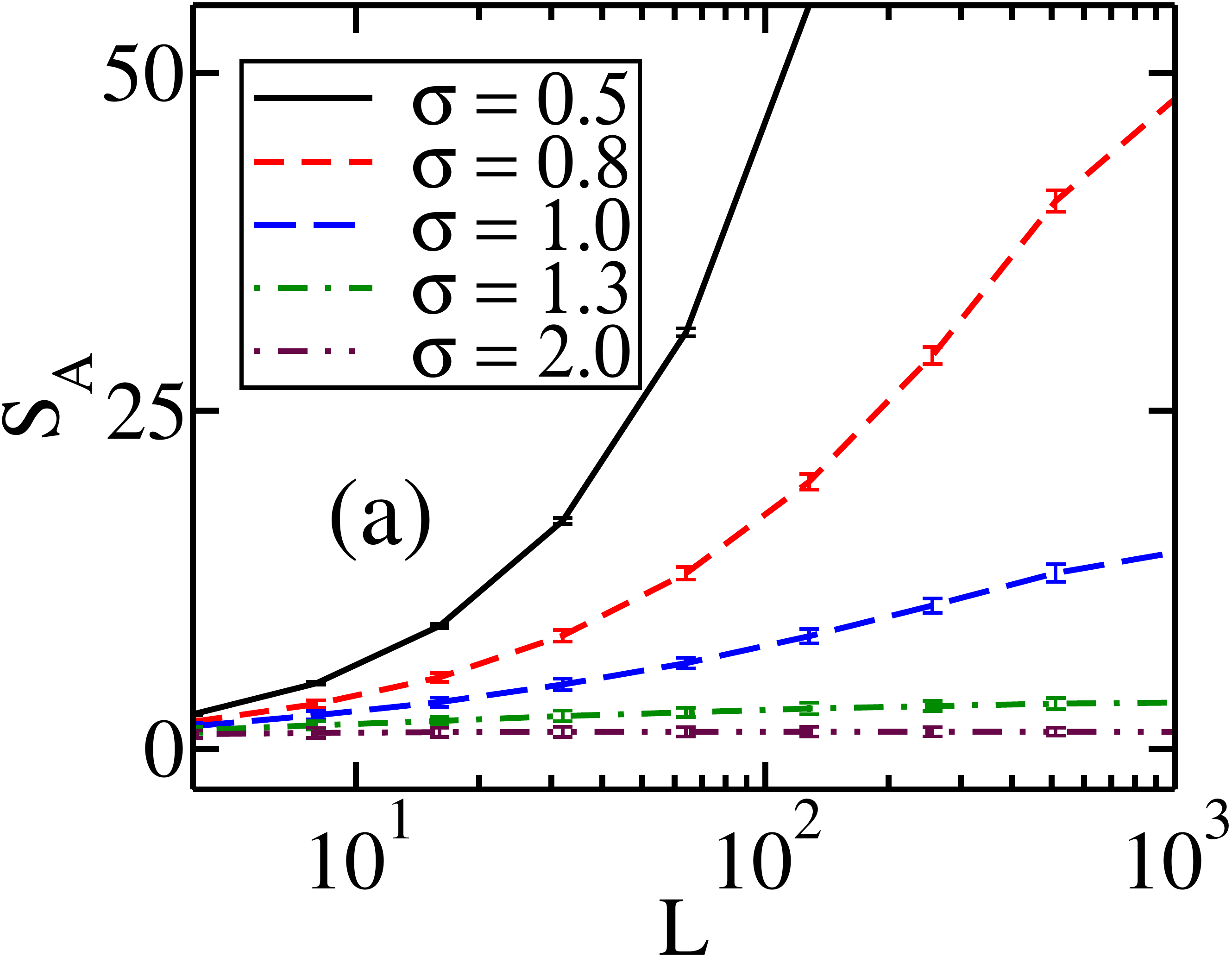}
\includegraphics[width=6.5 cm,height=5.0 cm]{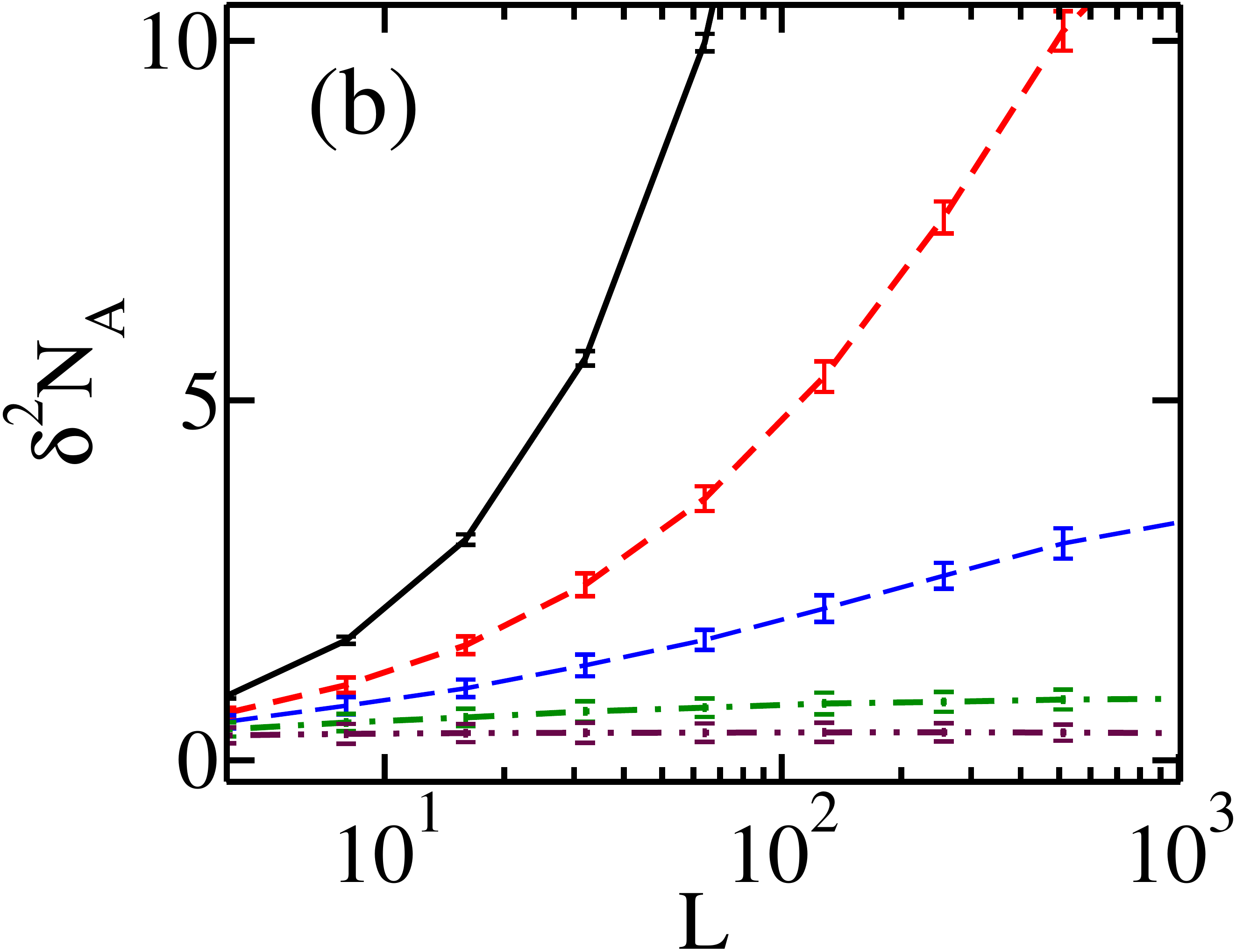}
\caption{(a) A linear-log plot showing the scaling of the entanglement entropy $S_A$ with subsystem size $L$ for increasing $\sigma$ for fermions at half-filling. (b) Similar plot for the number fluctuation $\delta^2 N_A$. For both the plots lattice size $N=2048$ and number of disorder realizations is $100$.}
\label{EE_scaling}
\end{figure}

Now we will discuss the scaling of the entanglement entropy with
subsystem size. Typically, short range models of noninteracting
fermions show logarithmic violation of the area law of entanglement
entropy i.e. $S_A \sim L^{d-1} \log L$ in $d$ dimensions~\cite{swingle}. 
In our disordered long-range model we see super-logarithmic
area law violation in the delocalized phase where $0<\sigma<1$. In fact it
goes as $L^\beta$, where the exponent $\beta=1$ at $\sigma=0$ and $\beta$ decreases as $\sigma$ increases
[Fig.~\ref{EE_scaling}](a). In the quasilocalized regime $1<\sigma<2$ it shows
area law for larger subsystem sizes whereas in the localized phase
$\sigma\geq2$ it shows a strict area law.

Next we discuss entanglement and its indirect experimental measurement. It has been argued~\cite{frerot} that fluctuations of a globally conserved quantity inside a subsystem can measure entanglement entropy as the quantity shares eigenfunctions with the reduced density matrix $\rho_A$ and hence provides a good basis for Schmidt decomposition of the many-particle eigenstate (see Ref.~\onlinecite{frerot} for a rigorous proof).
In our canonical set-up, total particle number is conserved and we study fluctuations in the particle number inside the subsystem, which is also an experimentally measurable quantity\cite{pitaevski,stringari}. The particle number fluctuations inside some subsystem $A$ can be defined as
\begin{equation}
\delta^{2}N_{A} = \sum_{i\in A}\langle n_{i}^{2}\rangle - \langle n_{i}\rangle^{2}.
\end{equation}
A close connection exists between entanglement entropy and fluctuations in the
local observables in the subsystem e.g. magnetization in a spin
system or particle number in free fermionic systems\cite{klich2009,song2010general,song2011,HFSong2012,calabrese2012exact,flindt2015}. The relationship
becomes a proportionality for certain gapless models, and the
proportionality constant to leading order has also been obtained~\cite{calabrese2012exact}. 

We adopt this quantity to study our long-ranged model, and look at the
scaling of the particle number fluctuations with the subsystem
size. The number fluctuations in the subsystem can be calculated using
the following relation:
\begin{equation}
\delta^2 N_A=\sum\limits_{m=1}^{L} \lambda_m(1 - \lambda_m).
\end{equation} 
Fig~\ref{EE_scaling}(b) reveals that this quantity shows exactly the same scaling as $S_A$, pointing to a proportionality between them, even in this long-range off-critical model. We will see that the \emph{proportionality constant} offers a signature for 
the localization-delocalization transition in the model though. Likewise, the proportionality constant shows a sudden jump at the phase transition in the AAH model as well, as will be shown at the end of this section. 

\subsection{Entanglement contour and fluctuation contour}
\label{sec:contour}
In this subsection, we will define and study the `entanglement contour'\cite{vidal}
and the `fluctuation contour'\cite{frerot}. These quantities contain microscopic
details of entanglement and number fluctuations. Specifically, the
contour keeps track of the contribution from each site within the
subsystem, to the quantity under consideration. Entanglement contour
is defined as the contribution ($C_s(i) \geq 0$) from the degrees of
freedom at each site $i$ in subsystem A to the entanglement entropy
$S_A$ such that $S_A = \sum\limits_{i \in A} C_s(i)$. One can calculate $C_s(i)$ using the
following relation\cite{vidal}:
\begin{equation}
C_s(i) = \sum\limits_{m=1}^{L} g_i(m) S_m,
\end{equation}
where $S_m = -[\lambda_m \log \lambda_m + (1 - \lambda_m) \log(1 - \lambda_m)]$. Here $\lambda_m$'s are the eigenvalues of the
correlation matrix $C$ or the entanglement spectra. $g_i(m)$ describes
the spatial pattern of the $m^{th}$ normalized eigenstate $\ket{\phi(m)}$ of matrix $C$ and hence of the entanglement Hamiltonian $H_A$
i.e. $g_i(m) = |\phi_i(m)|^2$. Similarly, one can define the
contour of subsystem particle-number fluctuations (also called as
`fluctuation contour') $C_n (i)=\langle \delta n_i \delta N_A
\rangle$, which is an obvious decomposition of the particle-number
fluctuations ($\delta^2 N_A$) in $A$ such that $\delta^2
N_A=\sum\limits_{i \in A} C_n (i)$. In the canonical ensemble $\delta
N_A=-\delta N_B$. Then $C_n (i) = - \langle \delta n_i \delta N_B \rangle$. 
So one can interpret $C_n (i)$ as the correlation between
number (density) fluctuations at site $i$ and those in the whole of
subsystem $B$. It can also be defined as\cite{frerot}
\begin{equation}
C_n(i) = \sum\limits_{m=1}^{L} g_i(m) \lambda_m (1 - \lambda_m),
\end{equation}
where all the terms have the same meaning as defined previously.
 
\begin{figure}
\centering
\includegraphics[width=7.0 cm,height=5.5 cm]{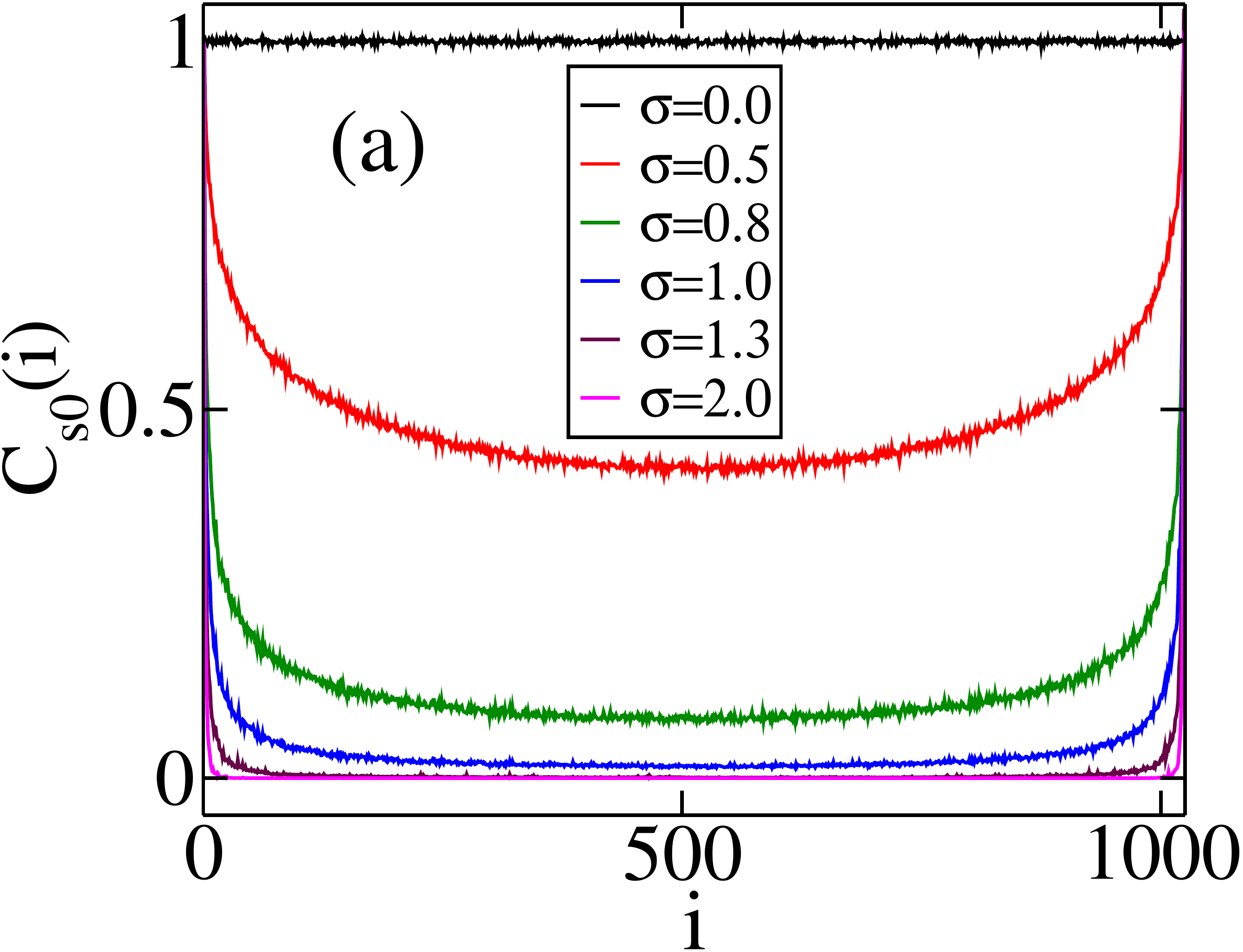}
\includegraphics[width=7.0 cm,height=5.5 cm]{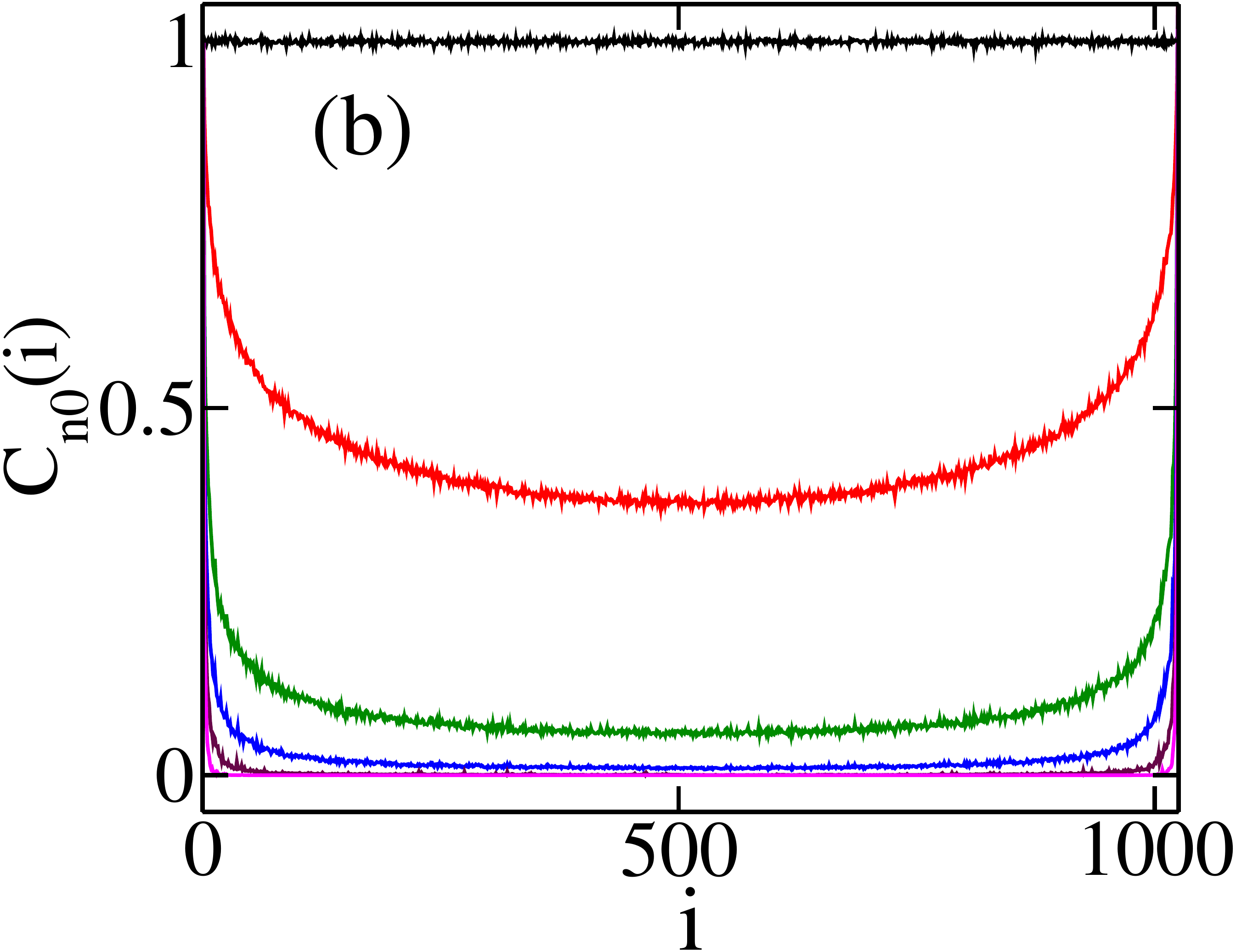}
\caption{(a) Spatial distribution of the scaled entanglement contour $C_{s0}(i)$ in the subsystem for different $\sigma$. (b) The same for the fluctuation contour $C_{n0}(i)$ for increasing values of $\sigma$.  For all the plots $N=2048$ and number of disorder realizations is $100$. Here $i$ is the site index in the subsystem and subsystem size $L=N/2$ for half-filled fermions.}
\label{contour}
\end{figure}
             
\begin{figure}[h!]
\includegraphics[width=7.5 cm,height=5.5 cm]{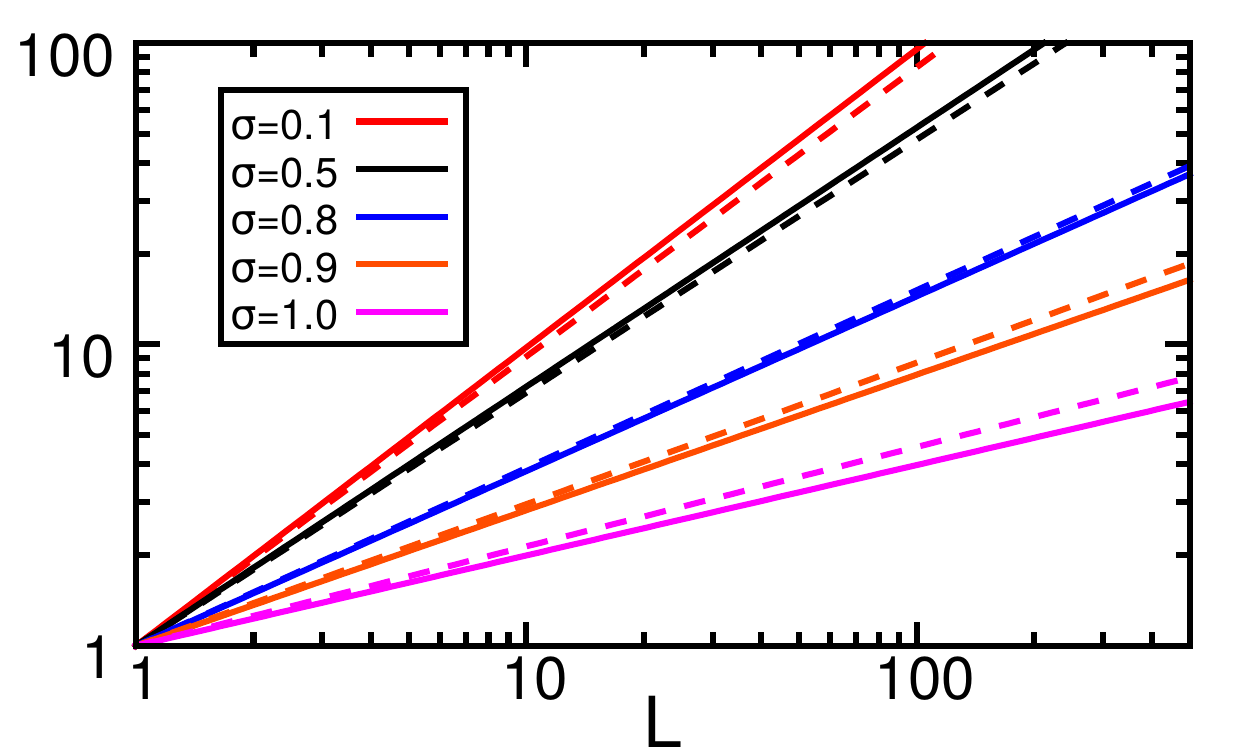}
\caption{The exponent $\beta$ calculated from the subsystem scaling of the entanglement entropy and the other exponent $\gamma$ determined from the decay of the entanglement contour in the subsystem are compared. In this log-log plot $L^\beta$ (solid lines) and $L^{1-\gamma}$ (dashed lines) are plotted to establish the relation $\beta\approx1-\gamma$ in the delocalized phase $\sigma<1$.}
\label{compare}
\end{figure}

It turns out that for free fermionic systems $C_n (i)$ and $C_s (i)$
show similar spatial dependence \cite{frerot}.  Spatial dependence of the $C_{s0}(i)=C_s(i)/C_s(1)$
entanglement contour and the scaled fluctuation contour $C_{n0}(i)=C_n(i)/C_n(1)$ of the random long-range
hopping model are shown in Fig.\ref{contour}(a) and in
Fig.\ref{contour}(b) respectively. Since there are two boundaries
between two subsystems in a ring and because the entanglement and the
number fluctuations decay as one moves away from the boundaries,
contours are symmetric functions of sites with respect to the
midpoint of subsystem $A$. We fit this decay with the function
$1/x^\gamma$. Since the entanglement entropy is the sum of all the
contributions of the entanglement contour, one may guess that the
entanglement entropy dependence should be given by the integral $\int
\frac{1}{x^{\gamma}}dx$, which in turn suggests that the exponent
$\beta$ should be given by $\beta \approx 1-\gamma$. Indeed, we find evidence
for this[Fig.~\ref{compare}], deep in the delocalized phase.

\begin{figure}
\centering
\includegraphics[width=4.2 cm,height=4.0 cm]{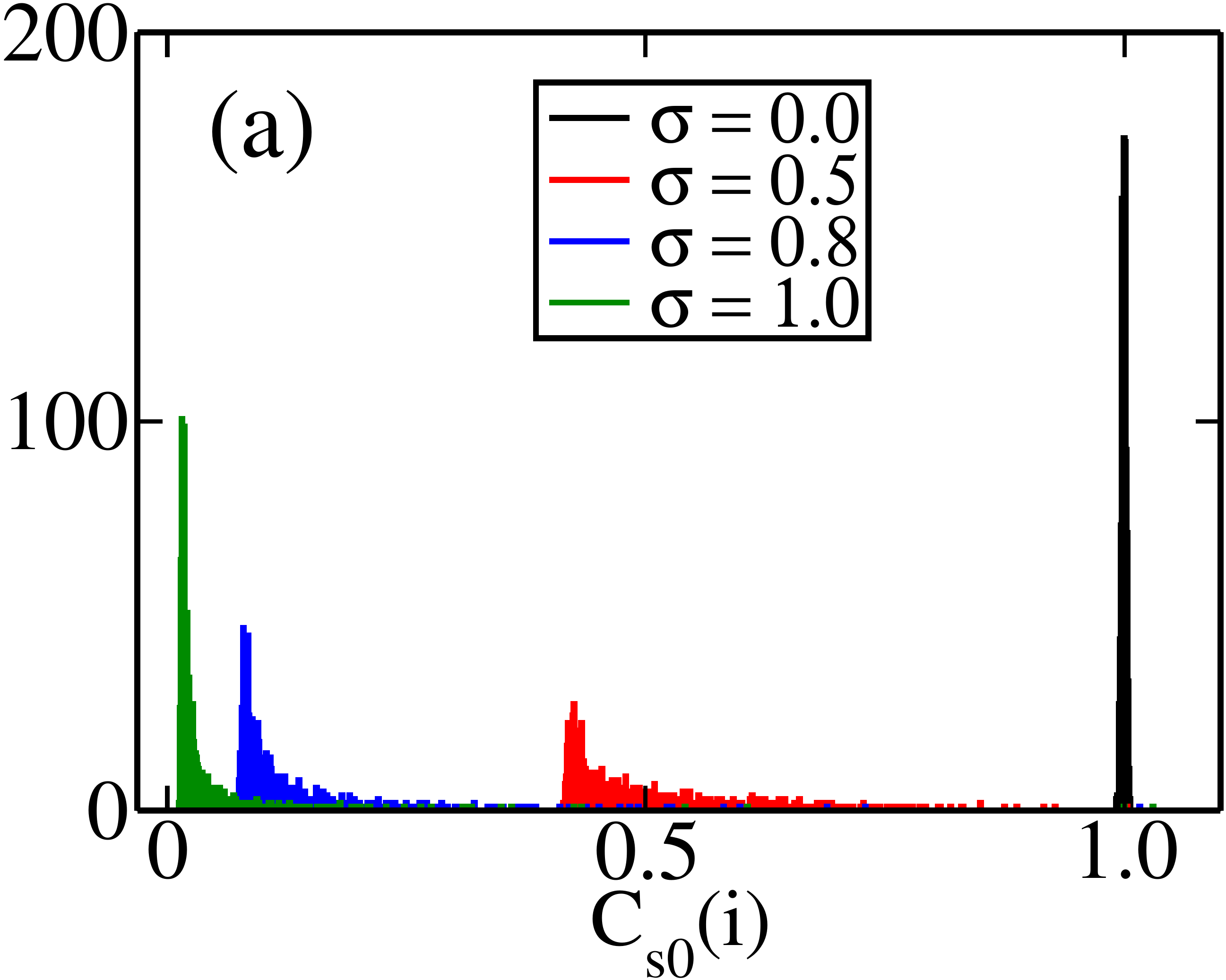}
\includegraphics[width=4.3 cm,height=4.0 cm]{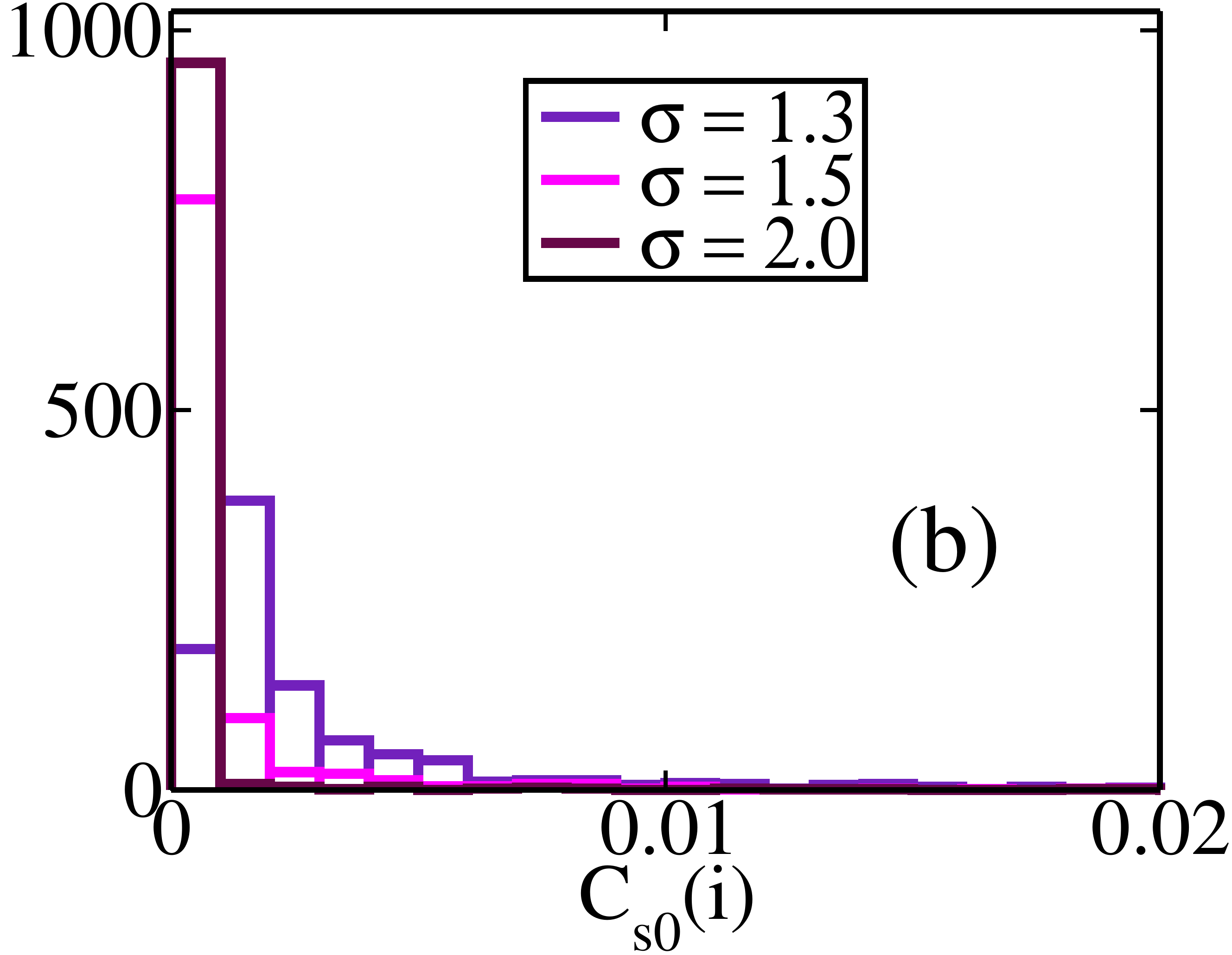}
\caption{ Histogram of the scaled entanglement contour $C_{s0}(i)$ (a) in the delocalized phase ($\sigma<1$); (b) in (quasi)localized phase ($\sigma>1$) respectively. For both the plots spinless fermions at half-filling are considered in a system of size $N=2048$. Here i indicates sites in the subsystem, whose size $L=N/2$ and number of disorder realizations is $100$.}
\label{histogram}
\end{figure}

For a finer understanding of the entanglement contour at the boundaries
and in the bulk of the subsystem, the histogram of $C_s(i)$ is plotted
in Fig.\ref{histogram}. In the delocalized regime, the entanglement contour has
a finite value at all the sites and the histogram is a sharply peaked distribution whereas the distribution gets broadened
and the peak shifts towards $0$ as one approaches the point of
quasi-localization $\sigma=1$ [Fig.\ref{histogram}(a)]. In the quasilocalized
regime the entanglement contour deep in the bulk starts vanishing [Fig.\ref{histogram}(b)], which explains the validity of
the area law for larger subsystem size. In the localized regime the
entanglement contour almost vanishes in the whole bulk region and one
gets a strict area law in this regime. This is also evident from the
histogram for $\sigma=2.0$ in Fig.\ref{histogram}(b) which shows a sharp peak at $0$ with
almost no broadening.
The fluctuation contour also shows similar behavior as the entanglement contour (not shown here).

\begin{figure}
\centering
\includegraphics[width=4.275cm,height=4.0cm]{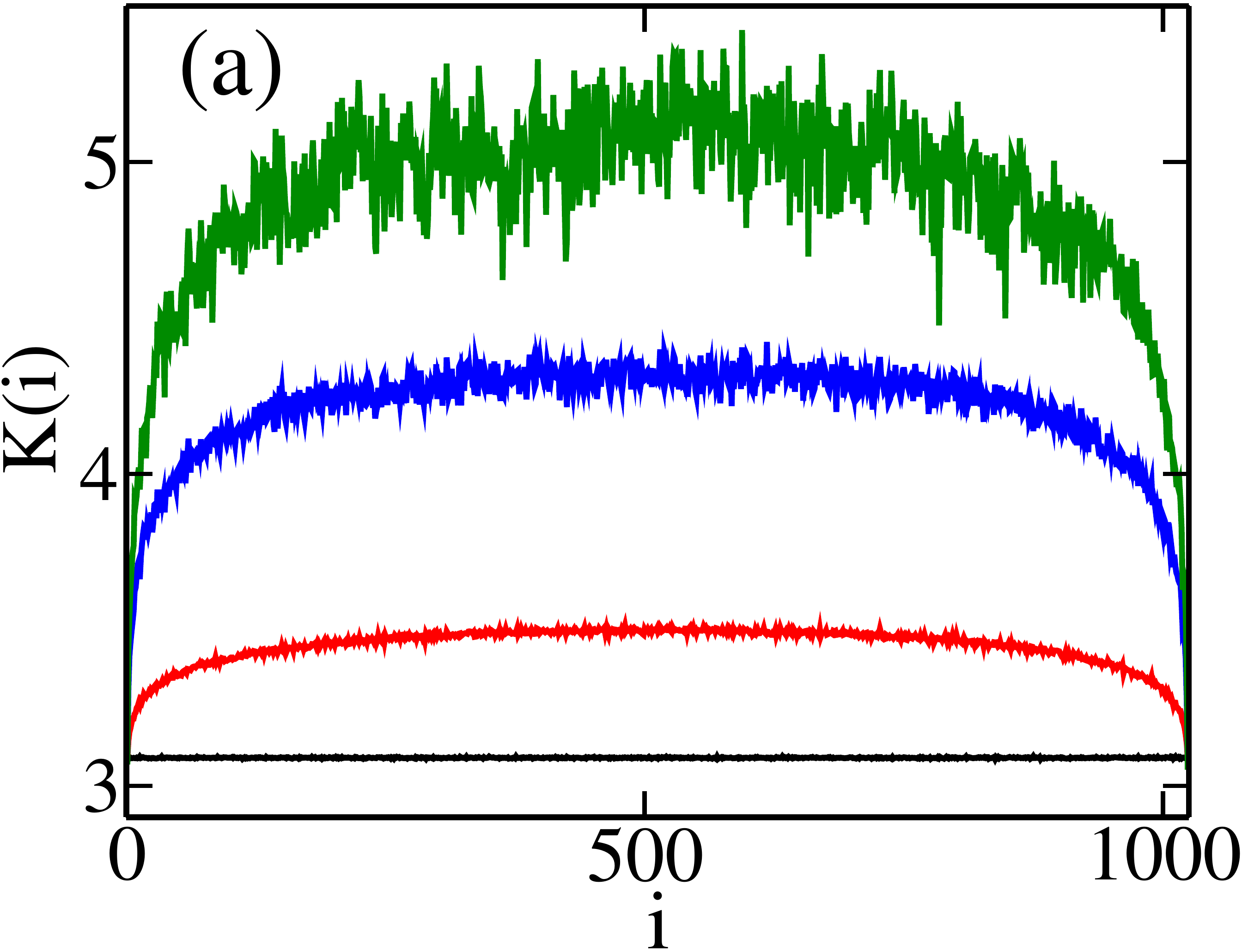}
\includegraphics[width=4.275cm,height=4.0cm]{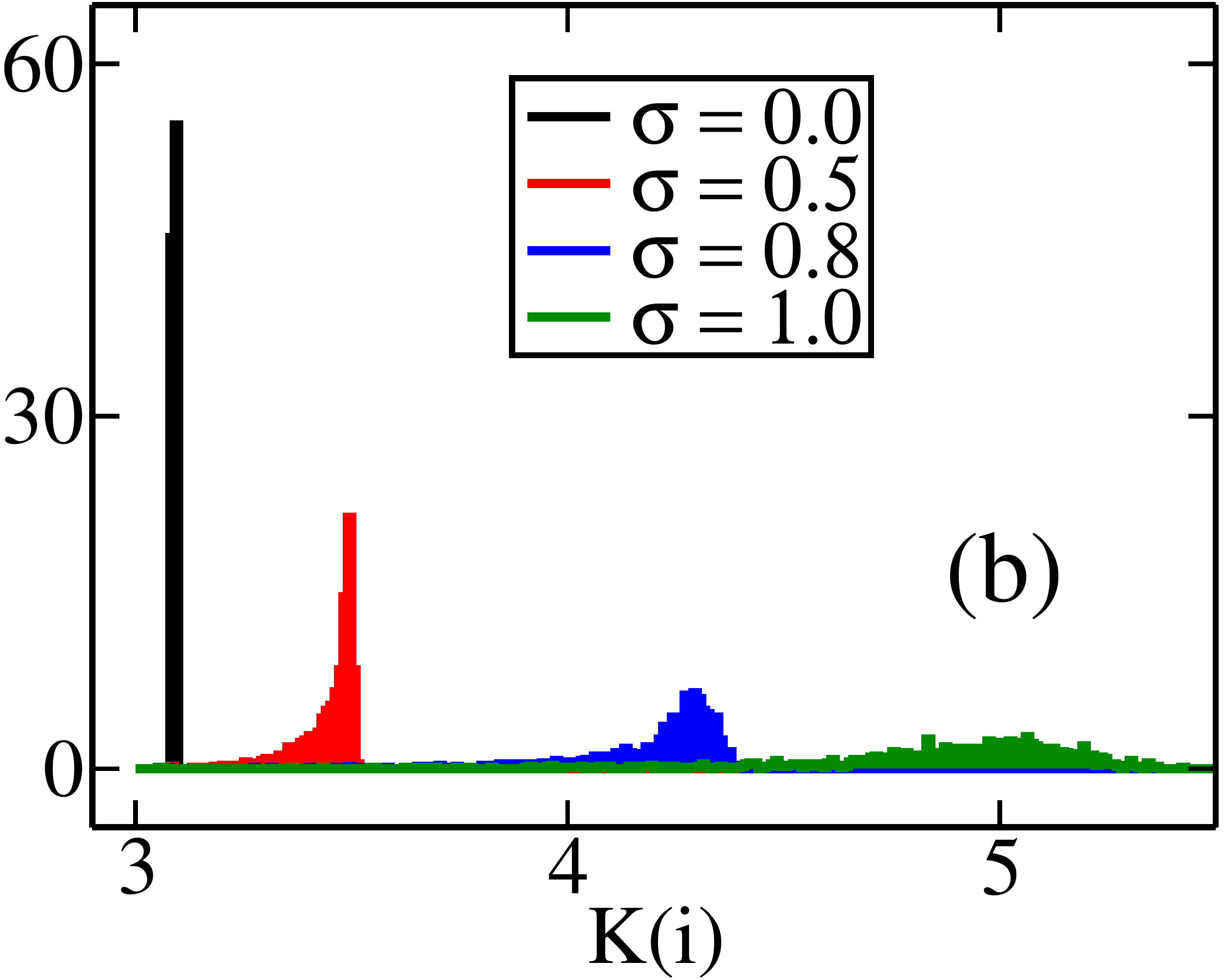}
\caption{(a) Spatial distribution of the ratio of two kinds of contour $K(i)$ of half-filled fermions for increasing $\sigma$ in the delocalized phase. (b) The corresponding histogram of $K(i)$. Here lattice size $N=2048$ and $i$ is the site index within subsystem $L=N/2$ and number of disorder realizations is $100$.}
\label{ratio1}
\end{figure}

Since the entanglement entropy and local number fluctuations are
intimately related, it is useful to study this relationship at a
microscopic level by calculating the ratio of the two contours of the
related quantities i.e. $K(i)=C_s(i)/C_n(i)$. This ratio for
increasing values of $\sigma$ in the delocalized phase is shown in
Fig.\ref{ratio1}(a). It reveals a uniform proportionality between the
two contours in the deep delocalized regime. The proportionality
becomes non-uniform as $\sigma$ approaches the transition point
$\sigma_c=1$. In the (quasi)localized regime this non-uniformity
becomes so much worse that we omit these data in the interest of
clarity. A histogram in Fig.\ref{ratio1}(b) shows a peaked
distribution of $K(i)$ for smaller $\sigma$ and the distribution gets
broadened with almost vanishing peak for larger $\sigma$.
\begin{figure}
\centering
\includegraphics[width=4.275 cm,height=4.0 cm]{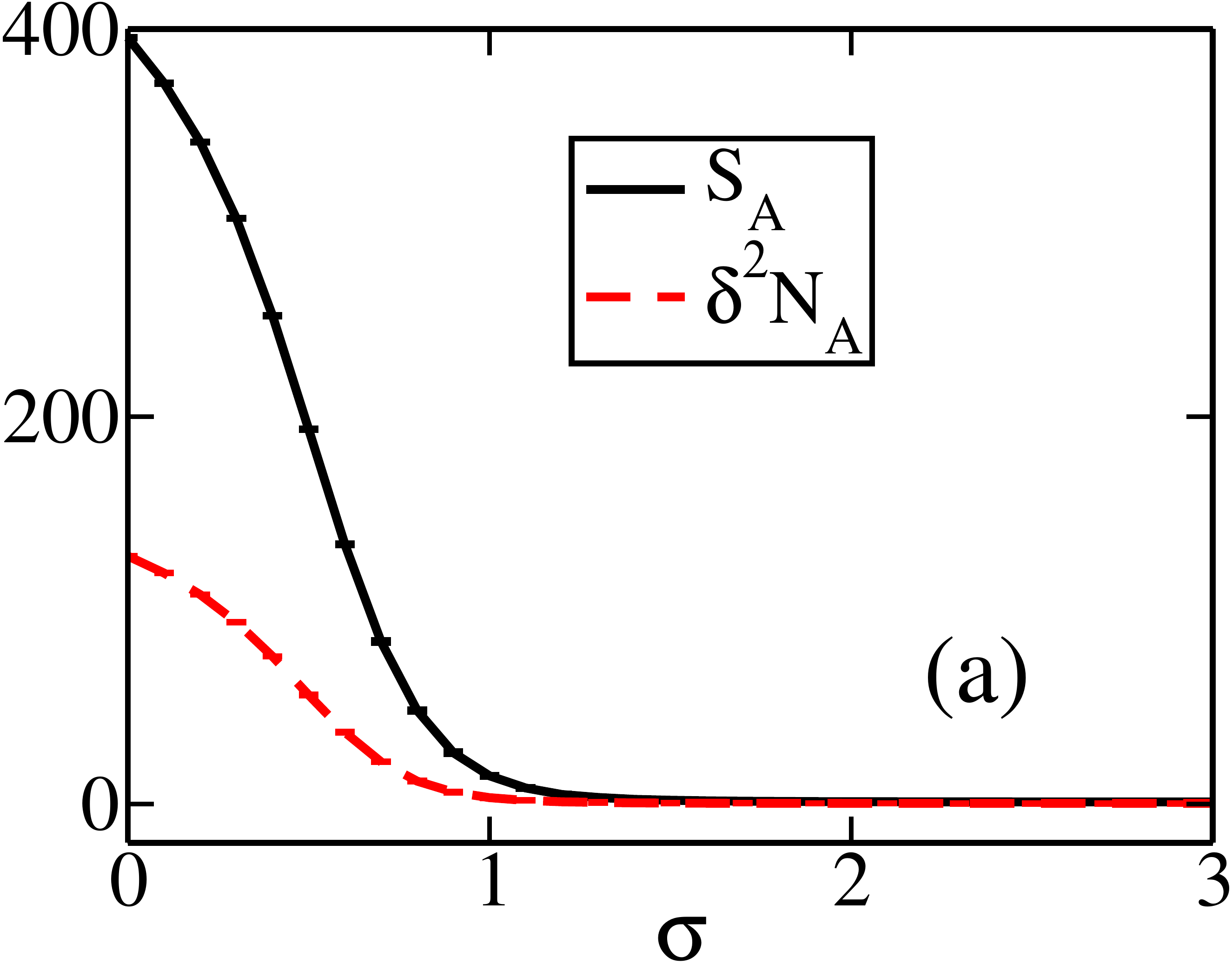}
\includegraphics[width=4.275 cm,height=4.0 cm]{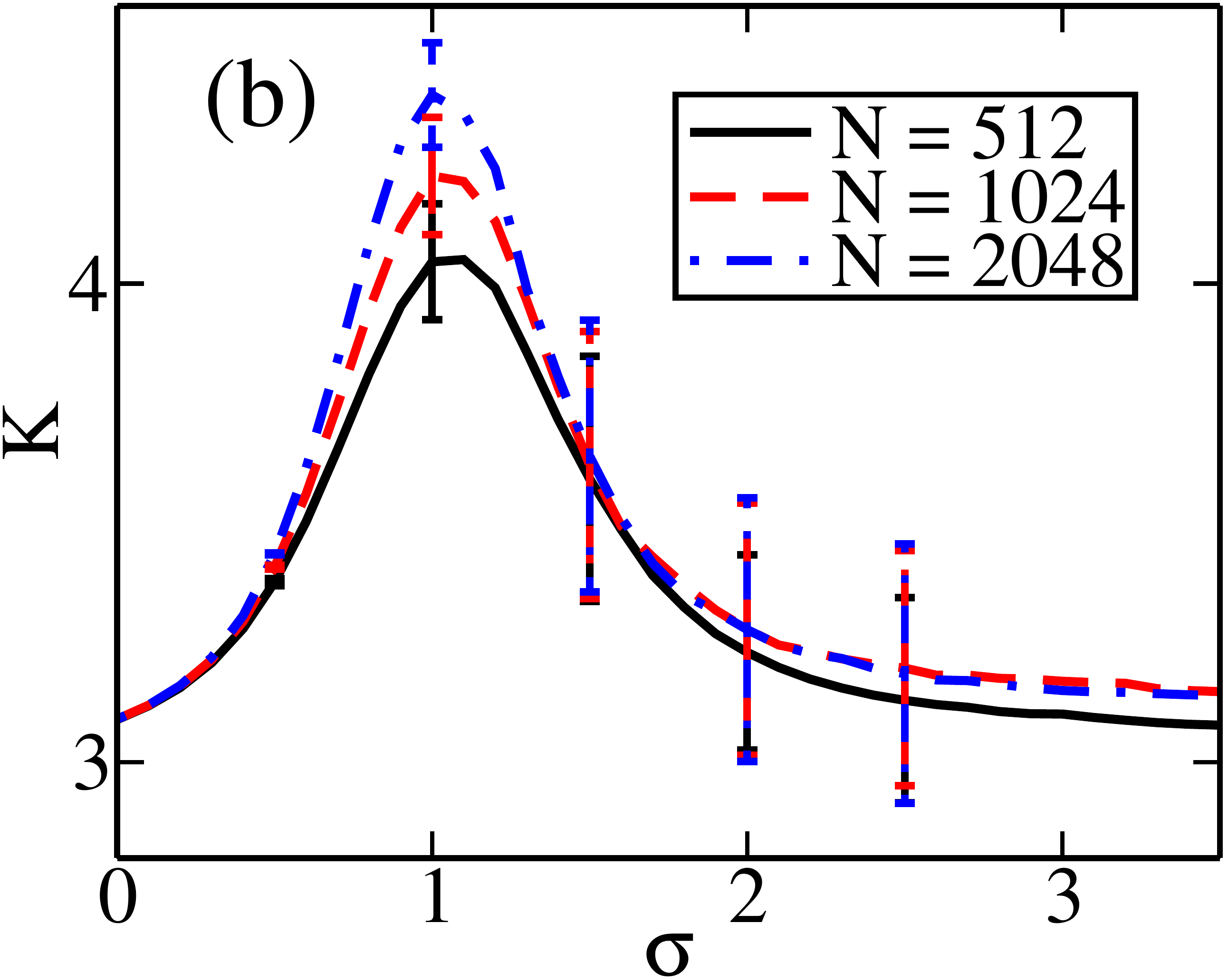}
\caption{(a) Variation of the entanglement entropy $S_A$ and number fluctuation in the subsystem $\delta^2 N_A$ of half-filled fermions with $\sigma$. Here the system size $N=2048$. (b) The ratio of the two quantities $K$ as a function of $\sigma$ for different system sizes $N$. In both the plots subsystem size $L=N/2$ and number of disorder realizations is $100$.}
\label{ratio2}
\end{figure}  
 
Next we study the proportionality constant $K$ of the relationship
$S_A = K \delta^2 N_A$ for free fermionic models. In a gapless system,
$S_A \propto \log L$ and $K=\pi^2/3$ for a 1D Fermi gas as shown in a
recent article\cite{calabrese2012exact}. However, in a gapped system
$K$ is not known in general; furthermore, $K$ is not believed to be a
universal quantity. This motivates us to investigate $K$ in our
long-range model. Though entanglement entropy and number fluctuations
in the subsystem vary in a similar fashion with $\sigma$
[Fig.\ref{ratio2}(a)], near the transition they conspire in such a way
that the ratio of them leaves a signature for the transition in the
model [Fig.\ref{ratio2}(b)]. The proportionality constant $K$ shows a
maximum at $\sigma_c=1$ and becomes almost constant in the localized
phase $(\sigma>2)$. Large error bars in the (quasi)localized regime in
Fig.\ref{ratio2}(b) are a reflection of the largely broadened
distribution of $K(i)$ in the same regime.
\subsection{Comparison with AAH model}
In the following, we have done a similar study as above in the AAH model which is a
short-range model that shows a sharp localization-delocalization
transition at finite disorder.
\begin{figure}
\centering
\includegraphics[width=4.275 cm,height=4.0 cm]{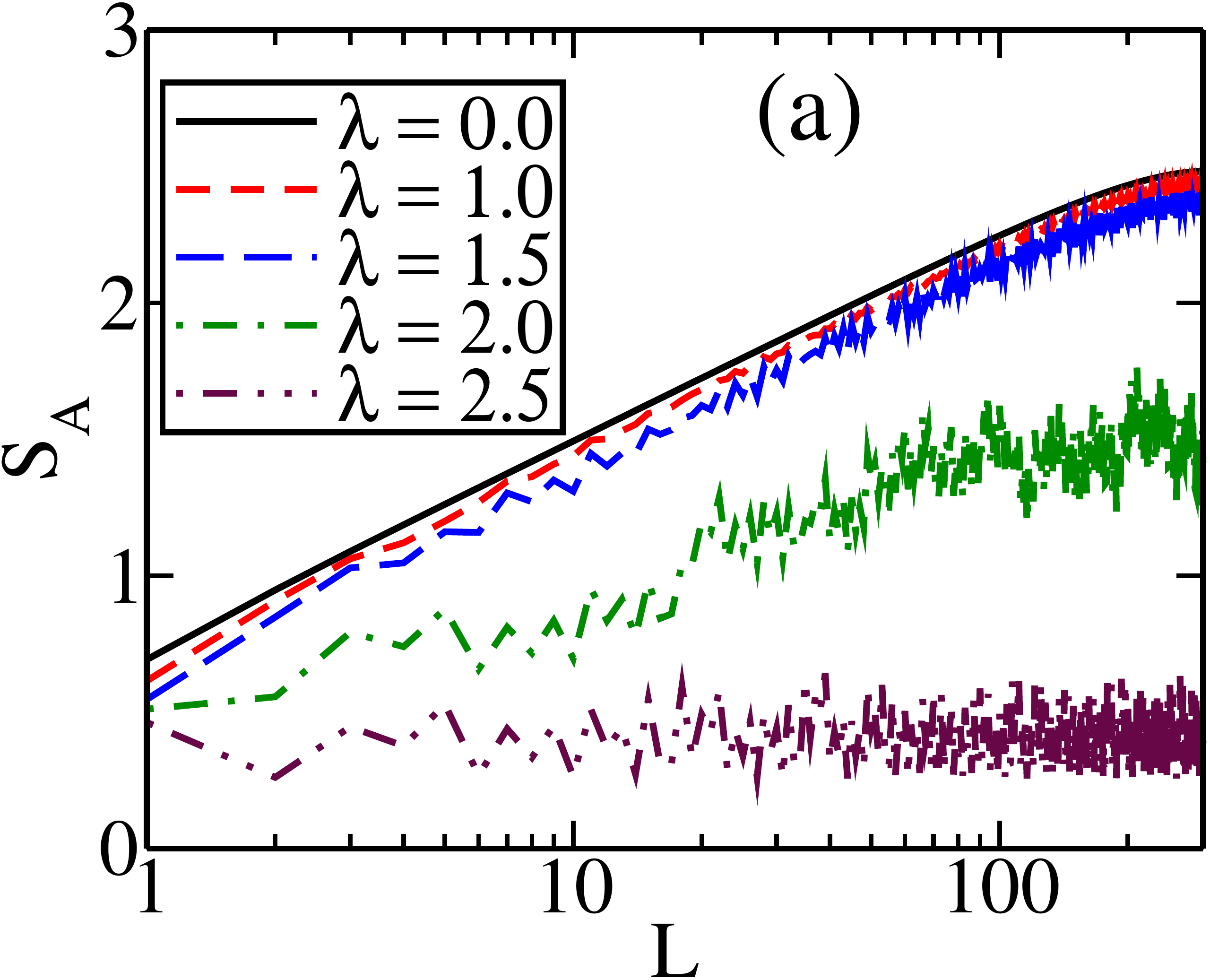}
\includegraphics[width=4.275 cm,height=4.0 cm]{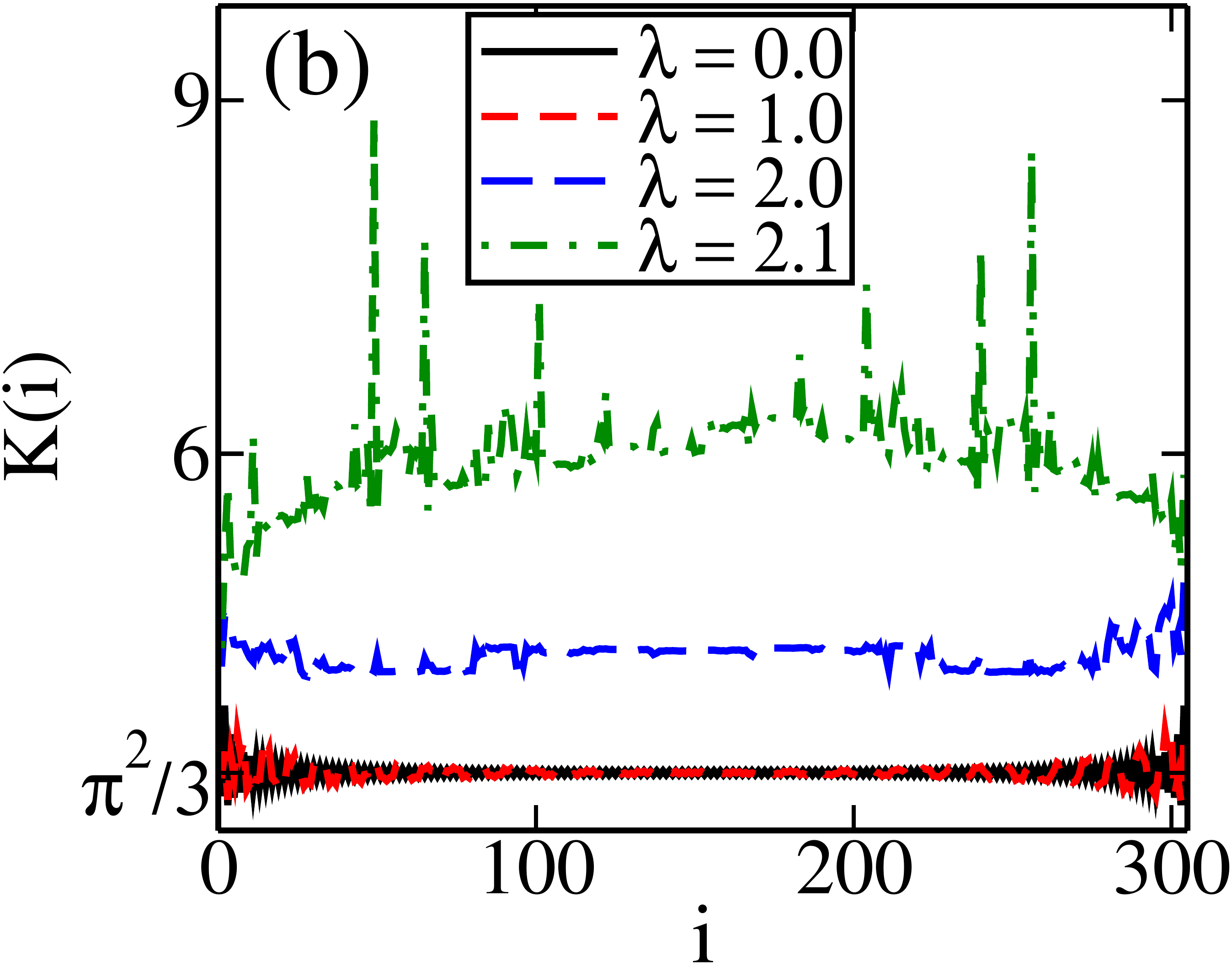}
\includegraphics[width=4.275 cm,height=4.0 cm]{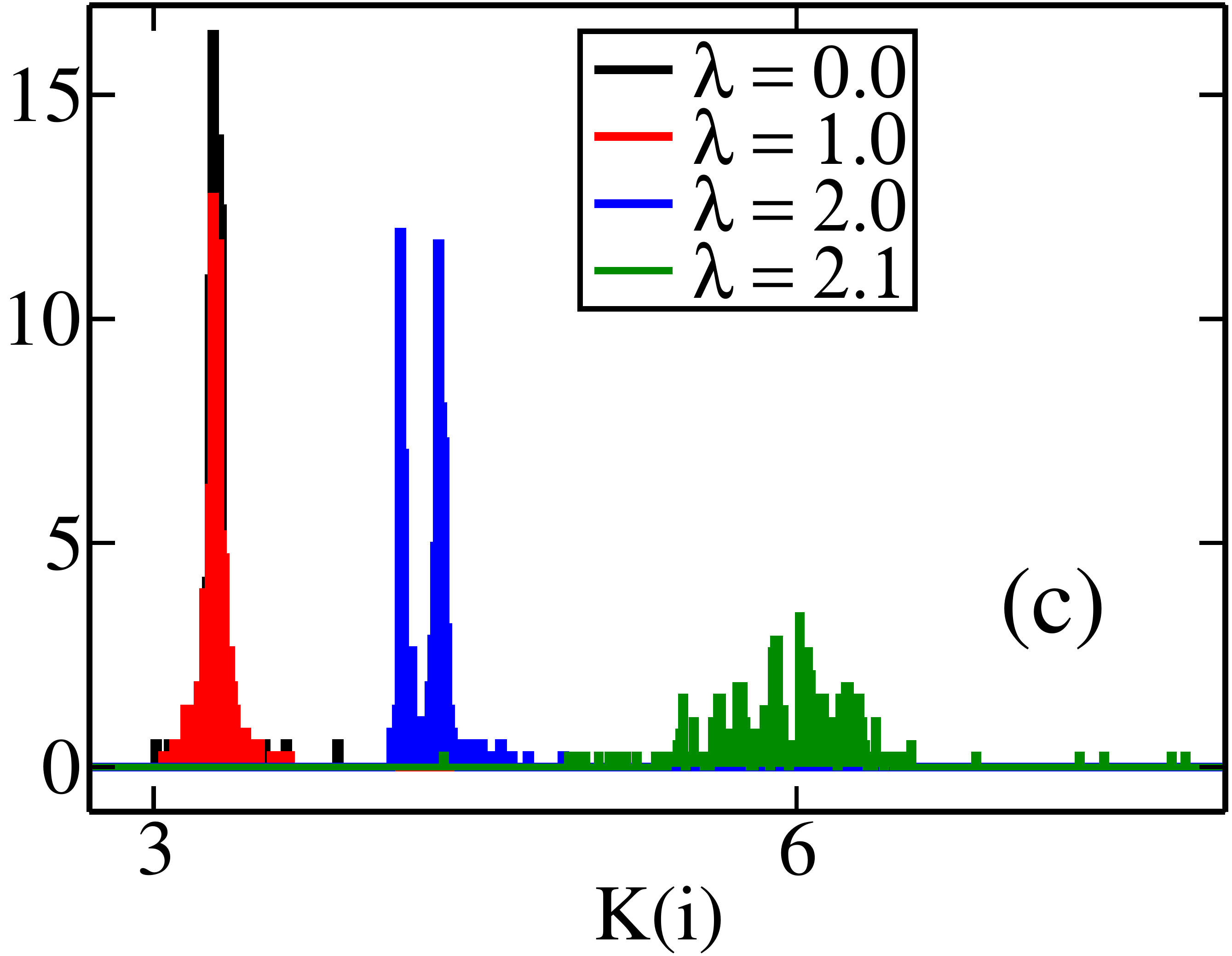}
\includegraphics[width=4.275 cm,height=4.0 cm]{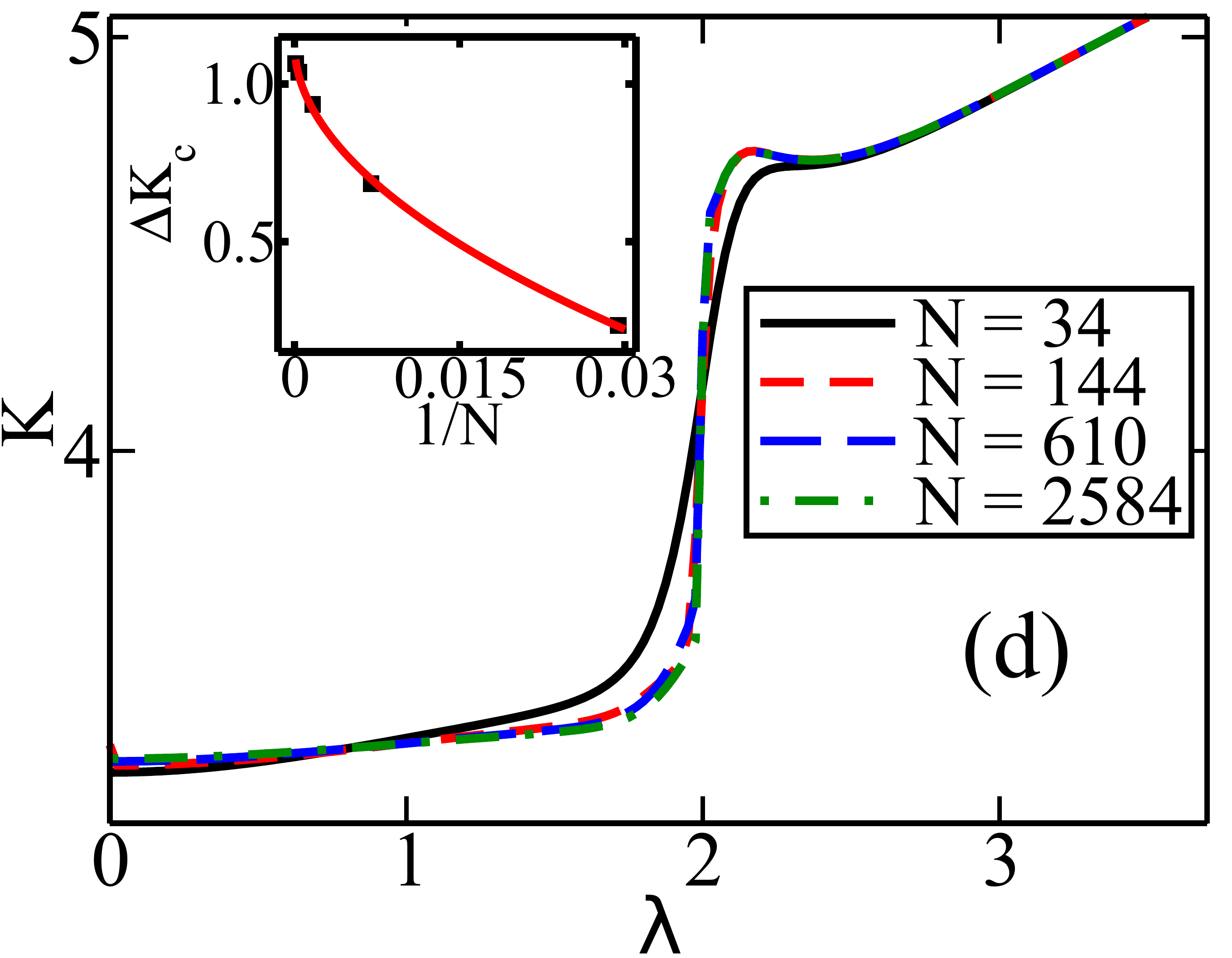}
\caption{Results from the 1D AAH model. (a) Scaling of the entanglement entropy $S_A$ with the subsystem size $L$ for increasing $\lambda$. The x axis is plotted in log-scale. (b) The ratio of entanglement contour to fluctuation contour $K(i)$ in the subsystem for different $\lambda$. (c) The corresponding histogram of $K(i)$. For the plots((a)-(c) $N=610$ (d) Proportionality constant $K$ as a function of $\lambda$ for increasing $N$. The inset is a fit to the $\Delta K_c$ vs $1/N$ data points(black square), where $\Delta K_c=K(\lambda=2.025)-K(\lambda=1.975)$. The red curve represents the fitting curve given by $\Delta K_c=1.12-5.47/N^{0.51}$. For all the plots, the subsystem size $L=N/2$ (except for figure (a)) for fermions at half-filling.}
\label{harper}
\end{figure}
The AAH model can be described by a
Hamiltonian of the same form as Eqn.\ref{hamiltonian} where
$t_{ij}=\delta_{i,j+1}$ and $v_i=\lambda \cos(2 \pi \eta
i)$. Here $\eta$ is a `Diophantine number' (e.g. $\frac{\sqrt{5} -
  1}{2}$, inverse of the `golden mean') and $\lambda$ is the strength
of the quasi-periodic disorder\cite{aubry,harper}. All the single particle eigenstates get
localized at $\lambda_c=2$\cite{mathieu}.

Our results for the Harper model are summarized in Fig.\ref{harper}.
In the tight binding model without any disorder $S_A \sim \log L$. As
the quasi-periodic disorder is turned on, in the delocalized phase
($\lambda<2$) $S_A$ retains the factor of $\log L$ which is a
modulated area-law behavior and in the localized phase ($\lambda \geq
2$) $S_A$ shows a strict area law, as shown in Fig.\ref{harper}(a).
In the delocalized regime $K(i)$ is close to $\pi^2/3$ in the bulk
whereas as one enters the localized regime it is no more a constant
and starts fluctuating [Fig.\ref{harper}(b)].  This is also evident
from the histogram of the same quantity. The distribution gets
broadened and the peak almost disappears in the localized phase
Fig.\ref{harper}(c). Also $K$ shows a jump at the transition point
$\lambda_c=2$ [Fig.\ref{harper}(d)]. We define $\Delta
  K_c=K_{\lambda=2+\delta \lambda}-K_{\lambda=2-\delta \lambda}$ near
  the quantum critical point $\lambda_c=2$, the scaling of which with
  the system size $N$ is well fitted by the functional form 
  $\Delta K_c=1.12-5.47/N^{0.51}$ for $\delta \lambda=0.025$ (the inset of
  Fig.\ref{harper}(d)). As $N\rightarrow \infty$, $\Delta
  K_c=1.12$. So when $\delta \lambda\rightarrow 0$, $dK/d\lambda$ will
  diverge to $\infty$ at $\lambda_c=2$ and hence the $K$ vs $\lambda$
  plot will become vertical at $\lambda_c=2$ in the thermodynamic
  limit. The proportionality constant $K$ indeed captures transitions
in the system although it changes differently in the two models
studied here.
\section{Non-equilibrium dynamics}
\begin{figure}
\centering
\includegraphics[width=4.275 cm,height=4.0 cm]{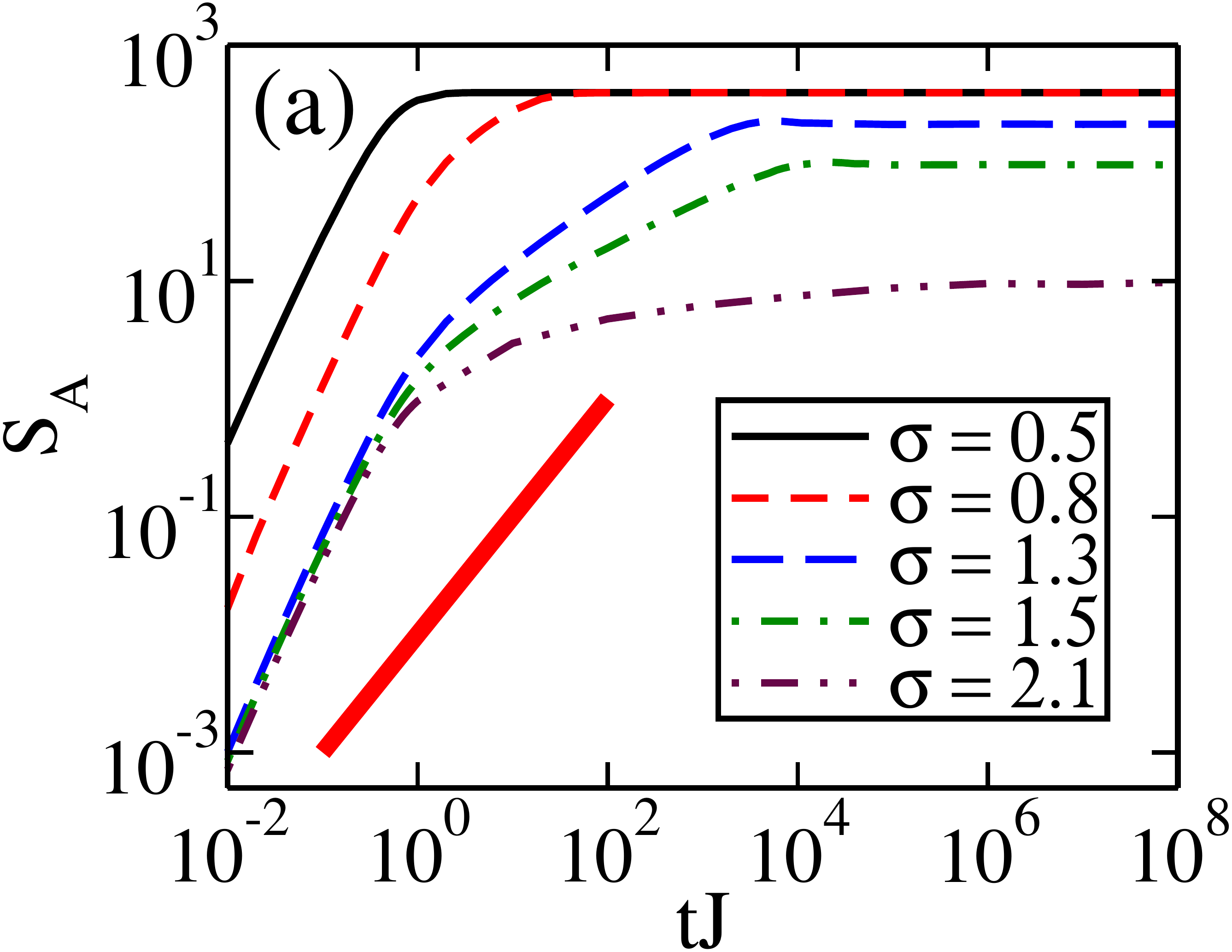}
\includegraphics[width=4.275 cm,height=4.0 cm]{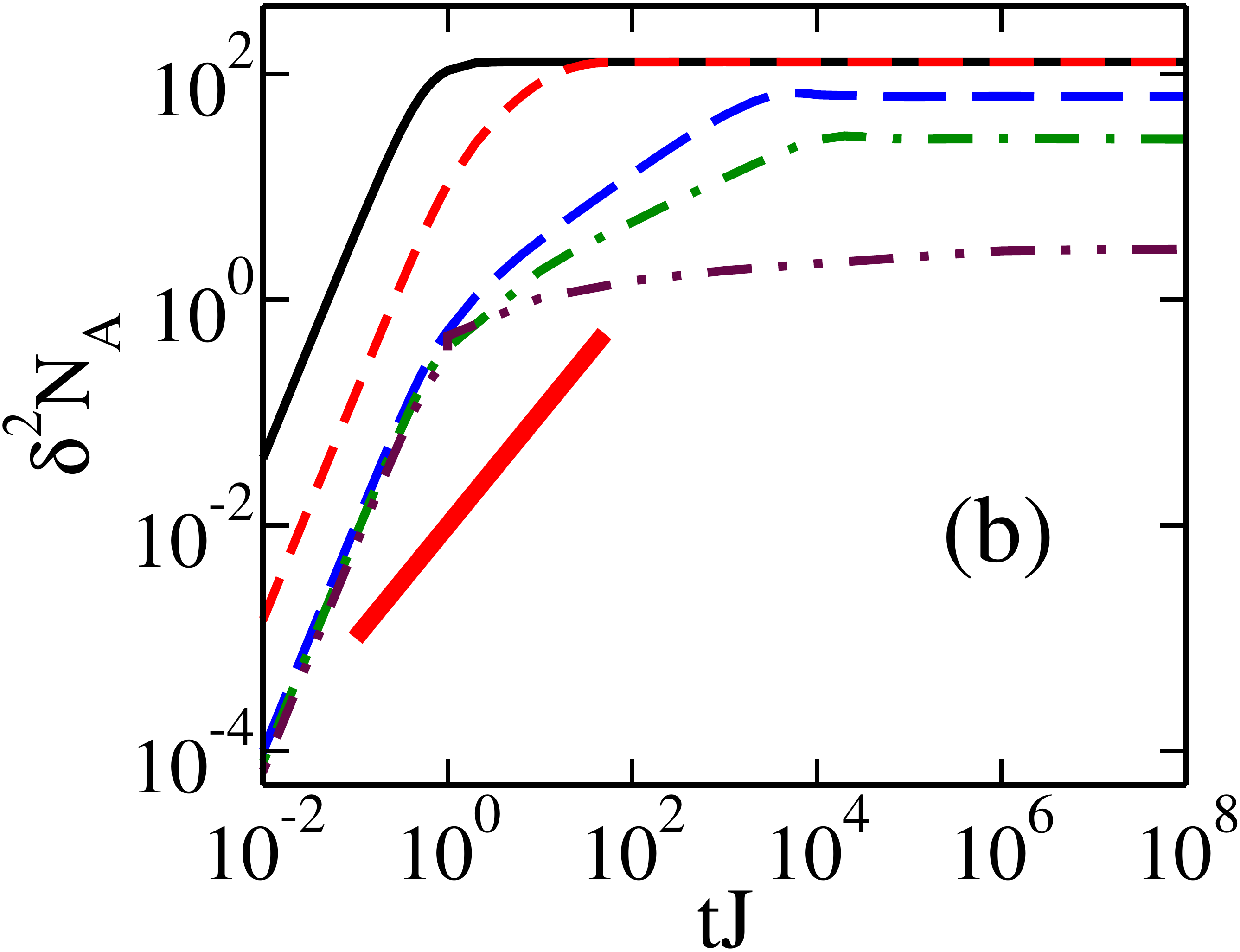}
\caption{(a) Quench dynamics of the entanglement entropy $S_A$ of half-filled fermions with time (in units of $J^{-1}$) for increasing $\sigma$ from an initial DW type state. (b) Similar plot for subsystem number fluctuations $\delta^2 N_A$. For both the log-log plots $N=2048$, $L=N/2$ and number of disorder realizations is $100$. The thick line segment shows linear dependence on time for comparison.}
\label{dw}
\end{figure}

Having studied the static quantities to analyze different phases, in
this section we investigate the dynamical properties of the model. A
non-equilibrium situation can be created by changing a parameter of
the Hamiltonian, locally or globally, through adiabatic or sudden
processes. Here we study the dynamics of entanglement entropy post a
sudden global quench in the bond-disordered long-range model and
compare the results with those of charge transport in the
system. We also briefly discuss correlation transport
  in the system in the context of velocity bounds on transport
  and the related light-cone picture~\cite{lieb}. We calculate the growth of
bipartite entanglement entropy $S_A(t)= -Tr(\rho_A(t)
\ln(\rho_A(t)))$, between two halves of the system A and B for our
model at half-filling. The data we present are with an initial state
of the density-wave(DW) type $\ket{\Psi}=\prod\limits_{i}
{c_{2i}}^\dagger \ket{0}$, which is evolved under the Hamiltonian at a
particular $\sigma$\cite{Fazio2006}. The DW state can be
  achieved by turning on an additional strong repulsive nearest neighbor interaction and
  then suddenly turning it off. We have checked that qualitatively similar results are
obtained when the initial state is the many-body ground state of
half-filled fermions corresponding to the Hamiltonian at $\sigma=2.5$,
with a quench carried out to various other values of $\sigma$.  To
calculate entanglement entropy, we use standard free fermion
techniques \cite{peschel2003calculation,Fazio2006}(see Appendix
\ref{appC} for details). Variation of $S_A(t)$ with time for the DW
type of initial state is shown in Fig.\ref{dw}(a). The entanglement
entropy varies with time in faster-than-linear fashion for $\sigma<1$
before it saturates, indicating the existence of a non-equilibrium
steady state. In the (quasi)localized regime ($\sigma>1$), after a
super-ballistic transient $S_A(t)$ goes in a sub-linear fashion with
time before reaching a saturating steady state. In the delocalized
phase, the saturation value $S_{A}^{\infty}$ barely changes with
$\sigma$; however in the quasi-localized phase, $S_{A}^{\infty}$
decreases with increasing $\sigma$. In the localized phase
($\sigma>2$) the entanglement growth becomes substantially suppressed as compared to the corresponding translationally
  invariant nearest neighbor model~\cite{cardy2005}, where the entanglement entropy
  reaches the saturation at a time $t_{sat}\sim L/2$. Also the
saturation values of $S_A$ in the localized phase are negligibly
small. The number fluctuations $\delta^2 N_A$, which are essentially
density-density correlations, reveal similar dynamics as $S_A(t)$
[Fig.\ref{dw}(b)].

\begin{figure*}
\stackunder{\includegraphics[width=5.5 cm,height=4.5 cm]{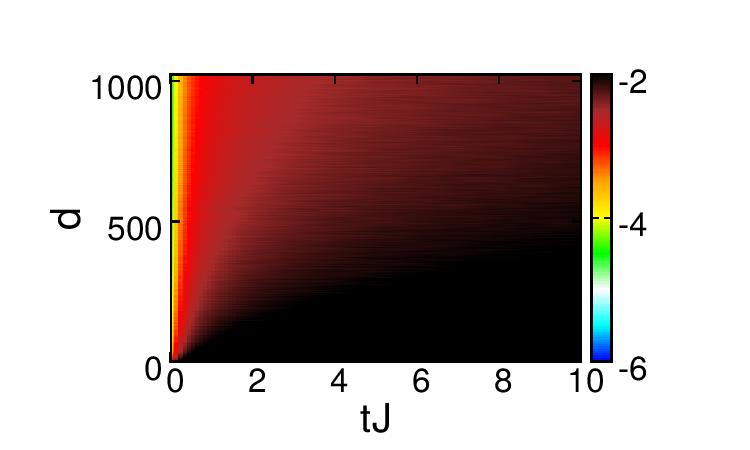}}{(a)}
\stackunder{\includegraphics[width=5.5 cm,height=4.5 cm]{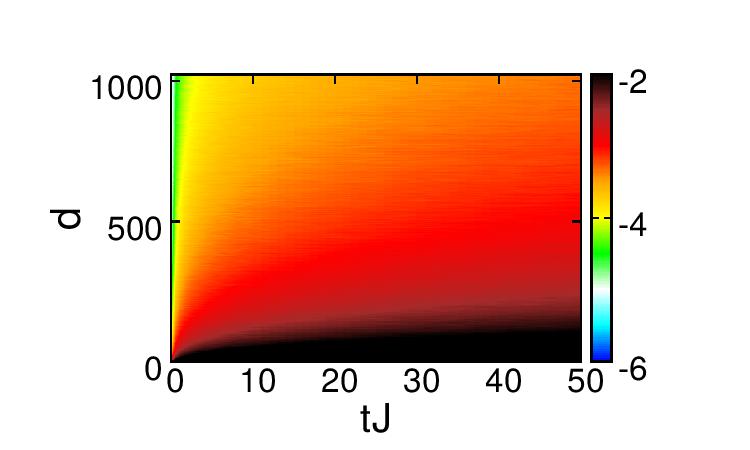}}{(b)}
\stackunder{\includegraphics[width=5.5 cm,height=4.5 cm]{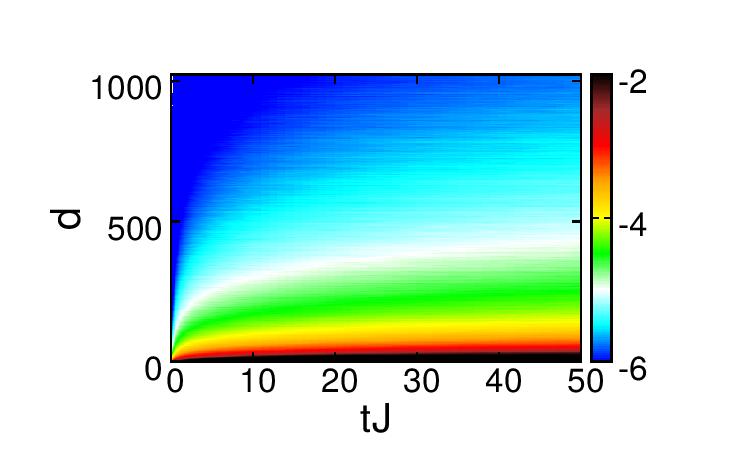}}{(c)}
\caption{(a)-(c) Surface plot showing the spreading of the amount of
  correlation $\log_{10}|C_d(t)|$ in time (in units of $J^{-1}$) and
  the lattice distance $d$ for $\sigma = 0.8, 1.3, 2.1$
  respectively. Here $d=|i-j|$, is the lattice distance between sites
  $i$ and $j$. The colors represent different ranges of values of the
  correlation, indicated in the bar legend attached to each
  figure - the numbers mentioned in the bar legend are the powers to be which ten is raised. 
  For all the plots half-filled fermions are studied with $N=2048$, $L=N/2$ and number of disorder realizations is $100$. The initial state is the DW state,
  described in the text.}
\label{corr}
\end{figure*} 

  In short-range models with translational invariance,
  following a global quench correlation transport happens with a
  constant velocity, defined as the Lieb-Robinson bound\cite{lieb},
  giving rise to a sharp causal light-cone like view of the
  correlation transport in space-time, outside of which correlations
  are exponentially suppressed\cite{cardy2006}. This leads to
  linear growth of entanglement entropy in such
  models\cite{cardy2005}. Breaking of translation invariance in
  short-range models can give rise to a much slower light cone, e.g. a
  logarithmic light-cone in the Anderson-localized phase\cite{burrell}
  and hence the entanglement entropy also shows a slow growth.  More
  than a linear growth of the entanglement entropy with time indicates
  the violation of the picture based on Lieb-Robinson bounds,
  which also bound the rate of growth of the entanglement. This kind
  of violation has actually been seen very recently in ultracold
  ionic experiments with translationally invariant long-range
  interacting spin models\cite{Gorshkov,zoller2014}. Also theoretical
  investigations have been carried out for translationally invariant
  long-range free fermionic models in this
  context\cite{kastner,daley,wouters}. To test the validity of the
  light-cone picture for correlation transport in our long-range
  free fermionic model with disordered hopping, we calculate the
  two-point correlation function $C_{d}=\langle{c_i}^\dagger
  c_j\rangle$ as function of time ($tJ$) and distance ($d=|i-j|$)
  between the sites $i$ and $j$ inside the subsystem as depicted in
  the surface plot in Fig.~\ref{corr}. At time $tJ=0$ the correlation
  matrix is diagonal with zero off-diagonal elements due to the
  product state structure of the initial DW state and the entanglement
  entropy is zero. At later times, different sites at distance
  $d=|i-j|$ start getting correlated. The correlation transport is
  more than linear or super-ballistic in nature within very short
  time-scales $tJ\sim 1$, which shows up as a transient in the
  quasi-localized $(1<\sigma<2)$ and localized $(\sigma>2)$ phases
  whereas in the delocalized phase, super-ballistic part is
  predominant as the time-scale for entanglement growth till it
  reaches the saturation is shorter ($t_{sat}J\sim 1$) in this
  case. This explains the super-ballistic entanglement growth in the
  system and violation of the picture based on Lieb-Robinson bounds. 
  However later time dynamics of the correlation reveals
  different behaviors of the light-cone picture in three different
  phases as we detail it in the following. As we can see from
  Fig.~\ref{corr}(a) for $\sigma=0.8$, one can still perceive 
  sub-linear light-cones in the delocalized regime. Sub-linearity
  indicates a decreasing velocity of the correlation transport with
  time as opposed to a constant velocity in the linear light-cone
  picture. The light-cone becomes more prominent and more sub-linear in
  the quasi-localized regime as can be seen in Fig.\ref{corr}(b). Very
  sharp sub-linear light-cones are visible in the localized regime
  (Fig.\ref{corr}(c)), where velocities of the correlation transport
  depend on the threshold values of the correlation. Sub-linearity of
  light-cones is more in this regime and hence the growth of entanglement
  entropy is very slow in the same regime. Such a change in light-cone picture from 
  less prominent to more prominent can be seen in the three regimes
  $(\sigma<1, 1<\sigma<2$ and $\sigma>2)$ also for the corresponding
  translationally invariant long-range hopping model with initial DW
  state\cite{kastner}. In contrast to our model though, in the non-disordered model, 
  all the light cones look linear and the related velocity bounds on the correlation 
  transport decrease as $\sigma$ decreases.

\begin{figure}
\centering
\includegraphics[width=4.275 cm,height=4.0 cm]{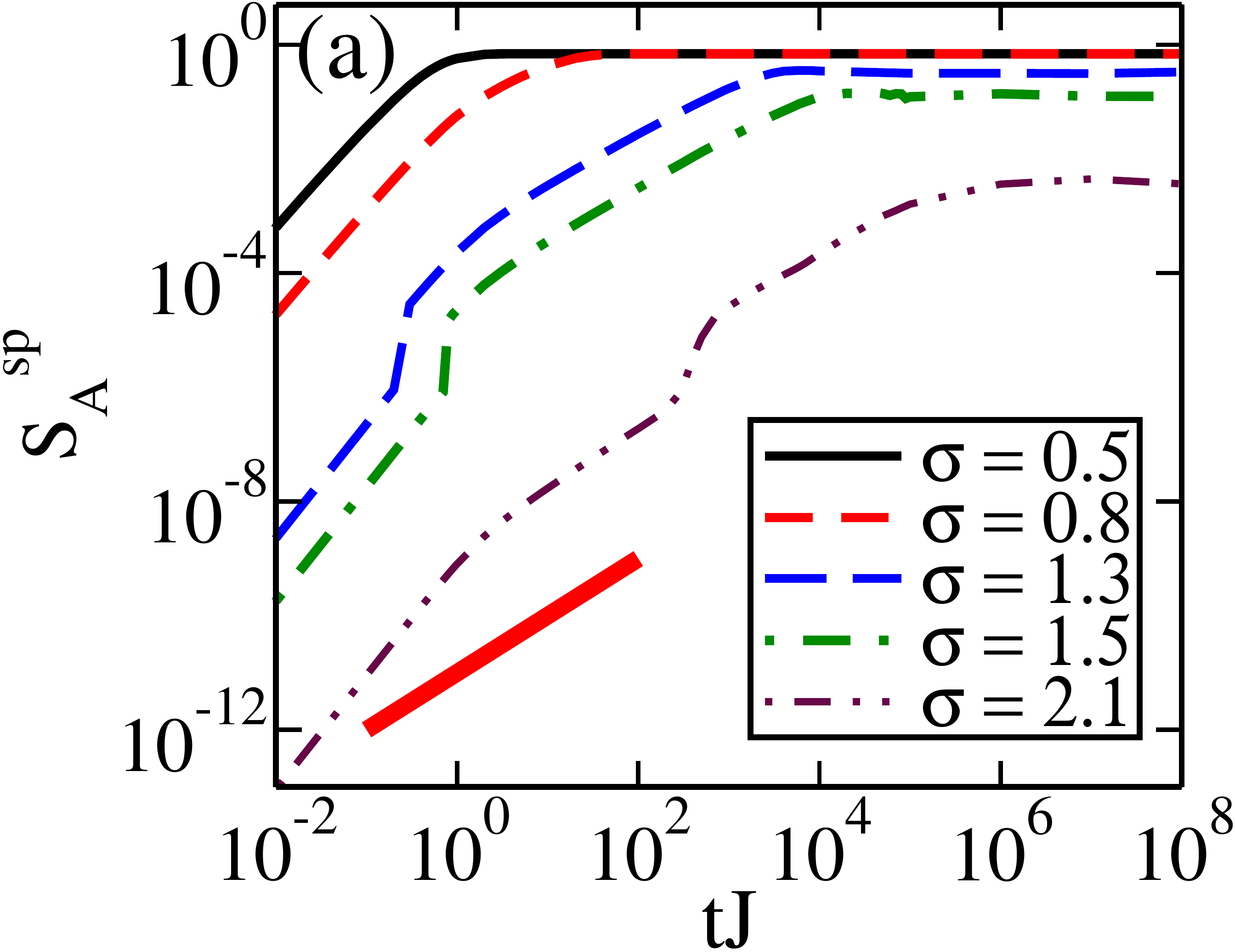}
\includegraphics[width=4.275 cm,height=4.0 cm]{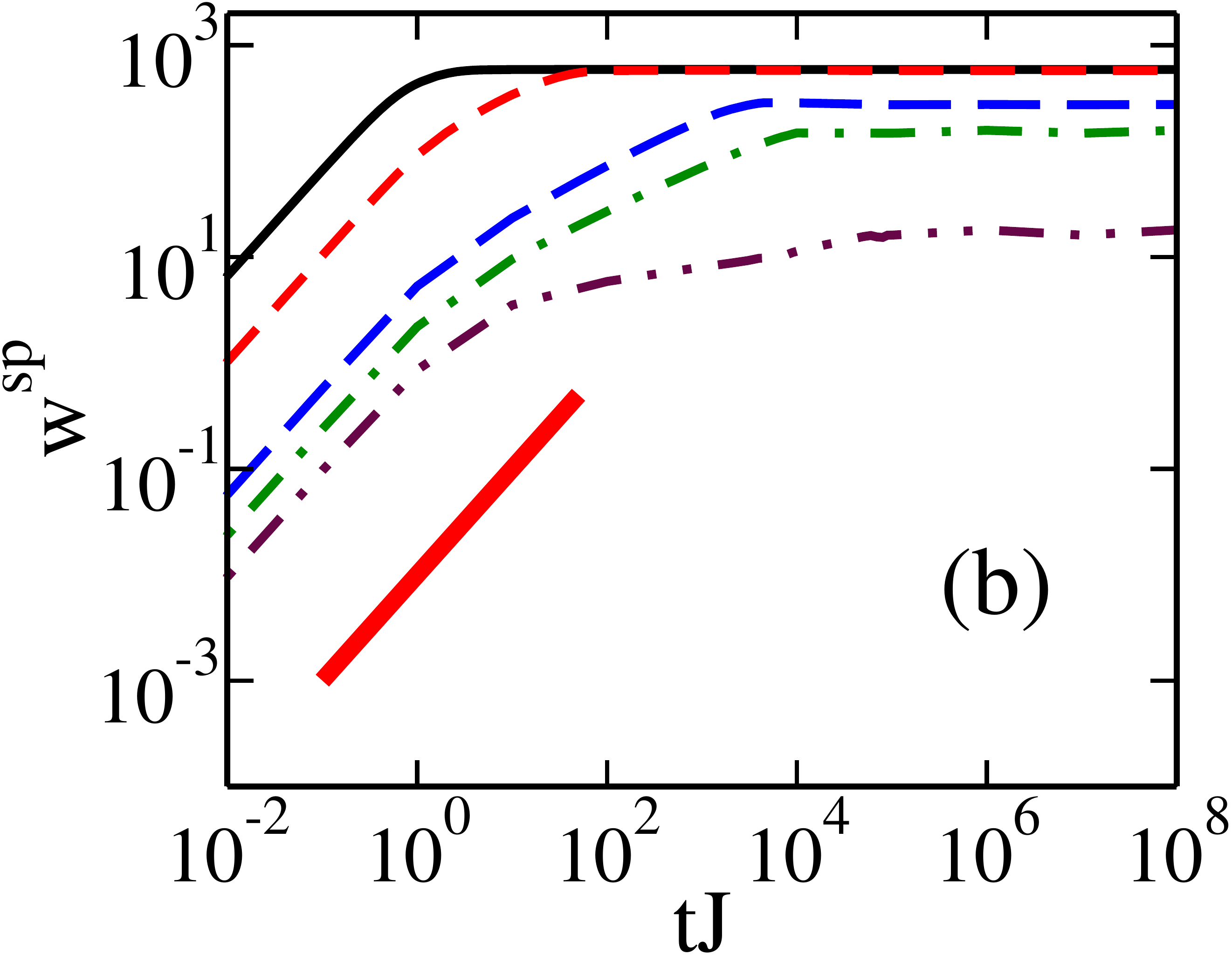}
\caption{(a) The single-particle entropy $S^{sp}_A$ as a function of time (in units of $J^{-1}$) for increasing $\sigma$. (b) Similar plot for the width $w^{sp}$ of the single particle wavepacket. The thick solid line shows the linear dependence on time for comparison. For all the log-log plots $N=2048$, $L=N/2$ and number of disorder realizations is $100$.}
\label{single}
\end{figure}

Next we will compare entanglement transport with charge transport in
the system at the single-particle and many-particle
levels. Single-particle entanglement entropy $S^{sp}_A$ is calculated
by choosing the subsystem $A$ such that it continues to be half the
size of the total system, but it is now taken to be centered around the
initial localized wavepacket in the middle of the lattice. The dynamics of $S^{sp}_A$ reveals
super-ballistic nature in the delocalized phase but in the
(quasi)localized phase the initial super-ballistic behavior is
followed by a ballistic part before saturation
[Fig.~\ref{single}(a)]. Also the average width of the initially
localized wavepacket is calculated, which is defined as
\begin{equation}
w^{sp}(t) = \sqrt{ \sum\limits_{i} (i-i_0)^2 p_i(t)},
\end{equation}
where $p_i(t)=|\psi_i(t)|^2$ as also mentioned earlier and $i_0$ is the center of the lattice. The dynamics of the width in
different phases is shown in Fig.\ref{single}(b).  In the (quasi)localized phase, after a ballistic
transient, $w^{sp}(t)$ goes sub-linearly before it reaches a saturation
value and the exponent of the sub-linear variation decreases as
$\sigma$ increases. This signals a sharp
contrast between charge transport and entanglement dynamics
even within the single-particle picture. Although both the quantities reach saturation at the same time, the saturation values decrease abruptly with $\sigma$ in the quasi-localized phase and becomes vanishingly small in the localized phase.
\begin{figure}
\centering
\includegraphics[width=4.275 cm,height=4.0 cm]{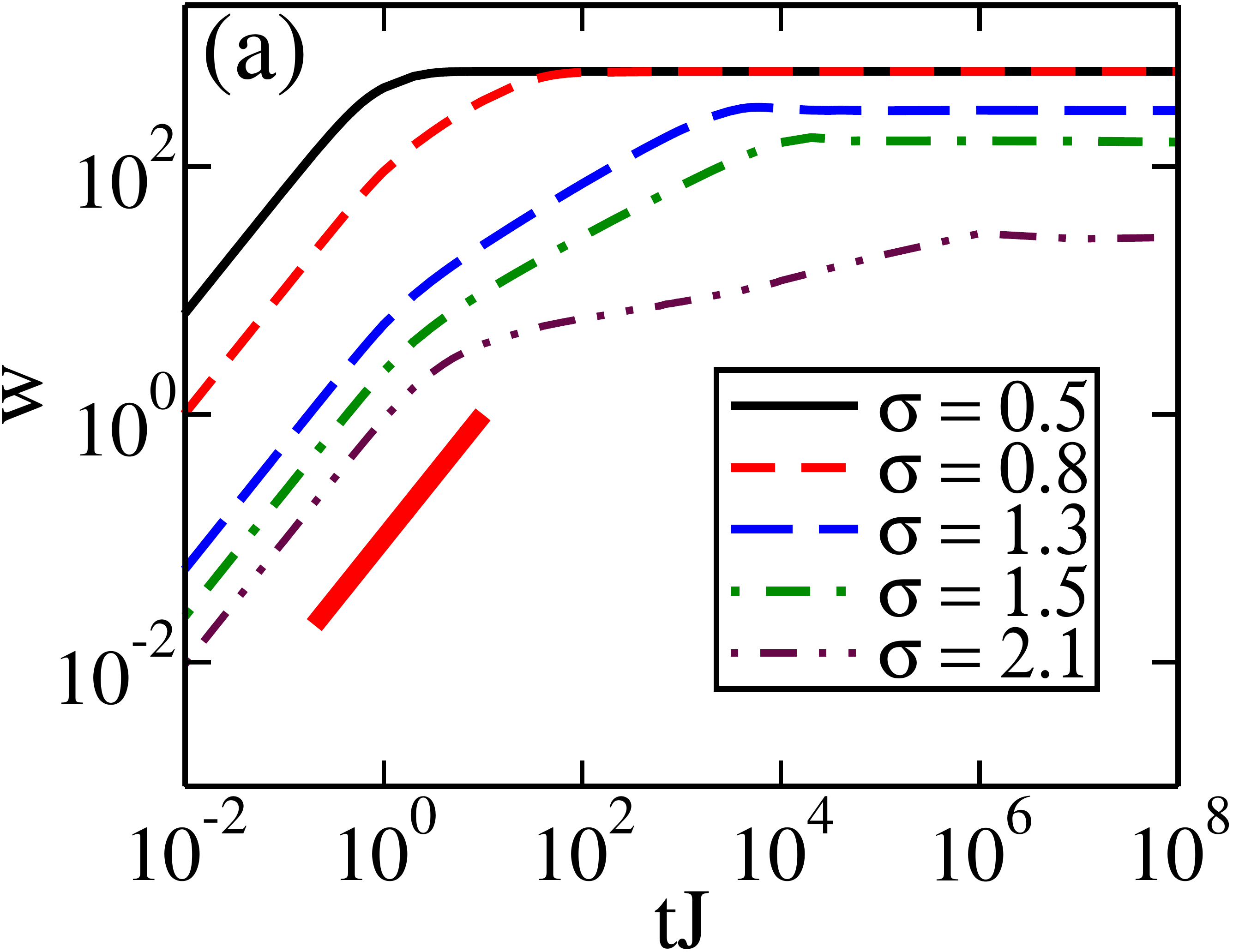}
\includegraphics[width=4.275 cm,height=4.0 cm]{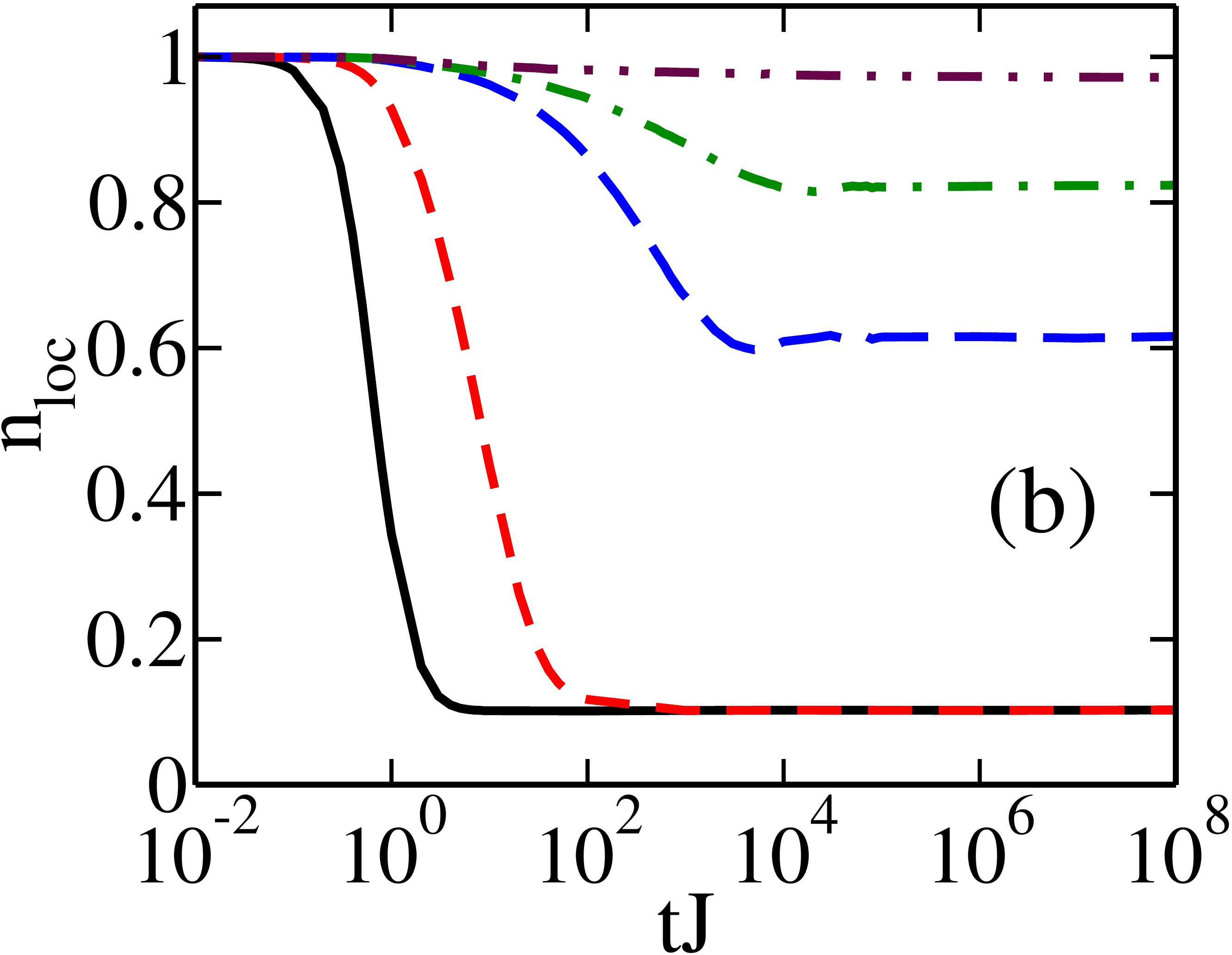}
\includegraphics[width=4.275 cm,height=4.0 cm]{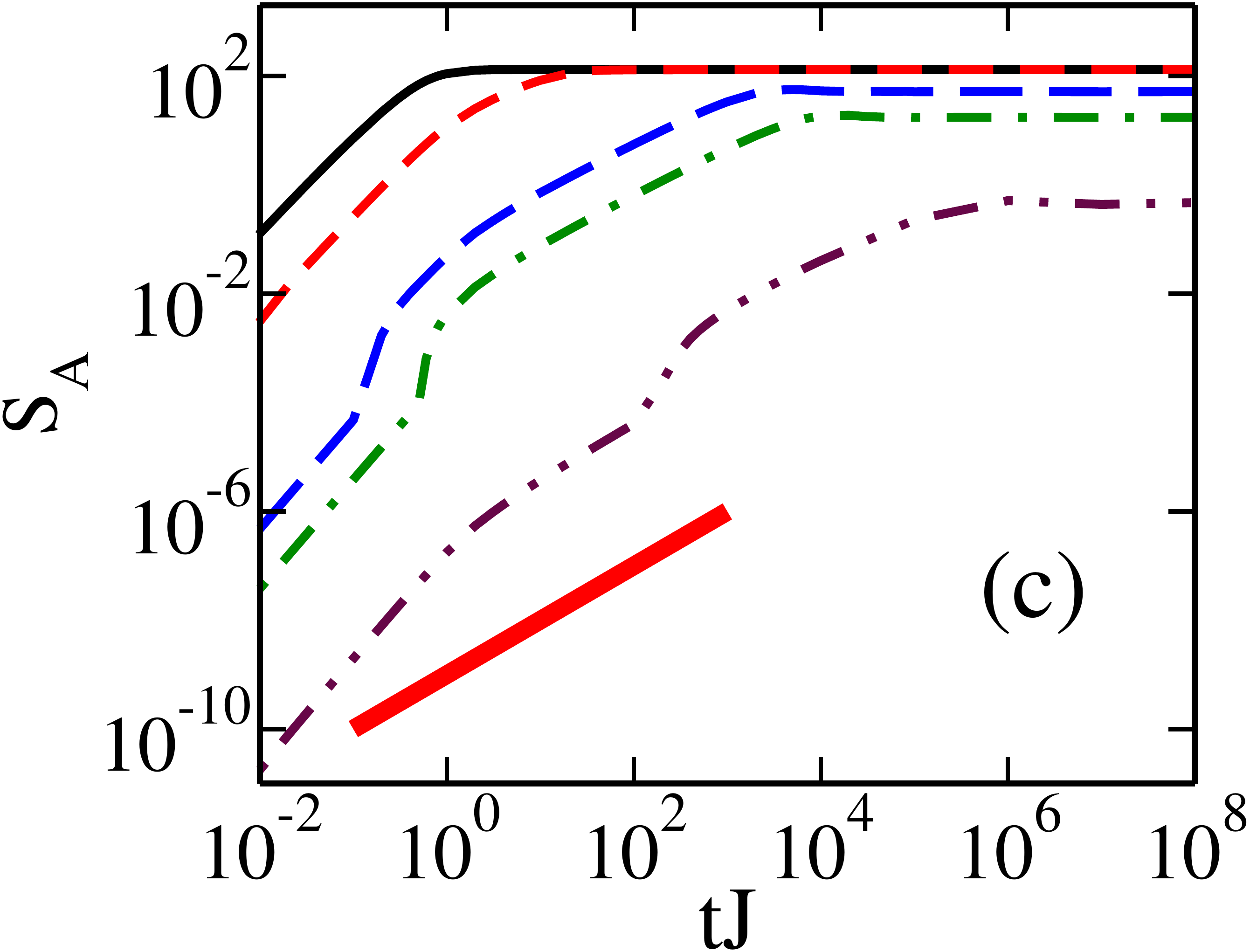}
\includegraphics[width=4.275 cm,height=4.0 cm]{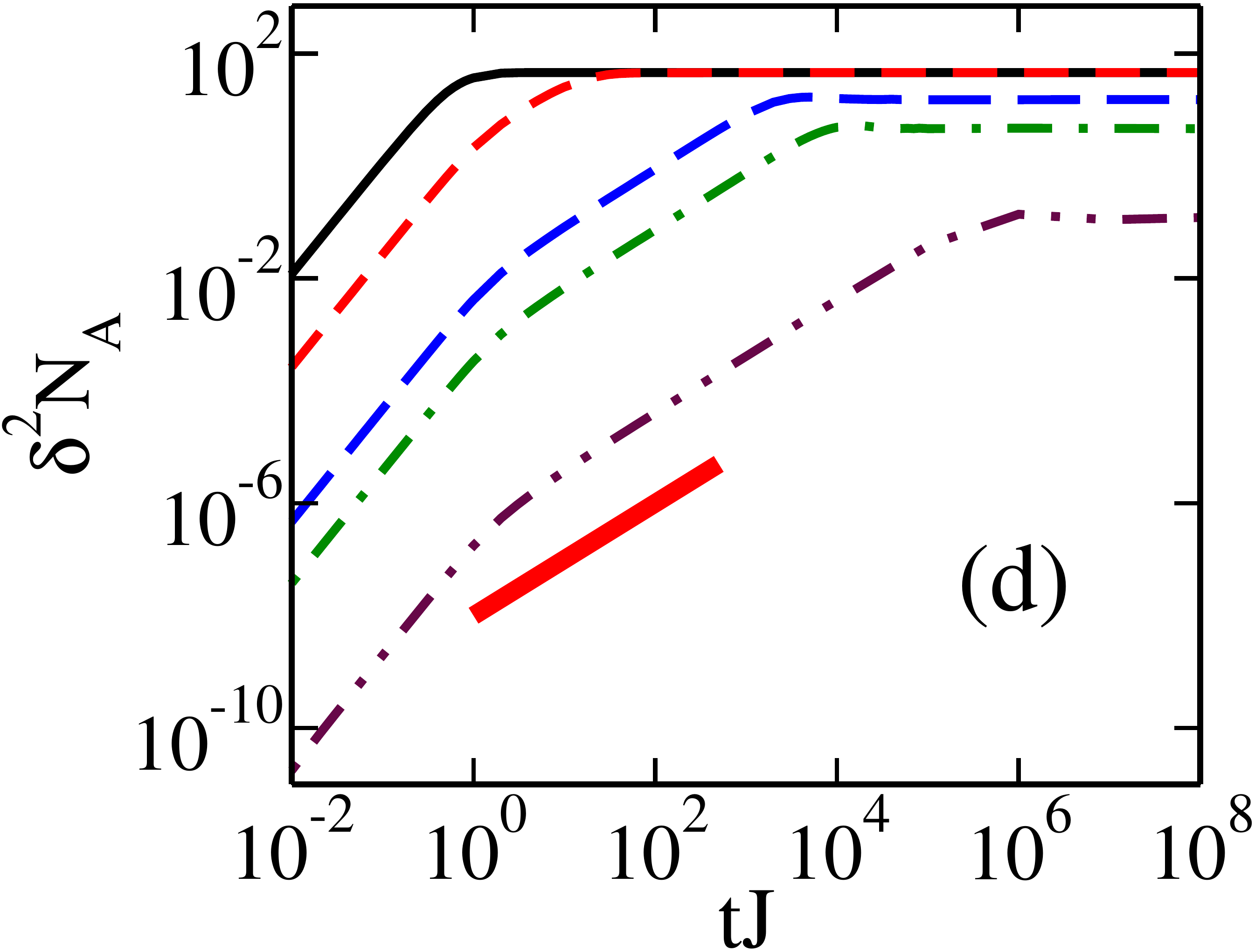}
\caption{(a)-(b) Variation of width $w$ of the many-particle wavepacket and the occupation density of the initially occupied sites $n_{loc}$ respectively with time (in units of $J^{-1}$) for increasing $\sigma$. In the delocalized phase $n_{loc}$ is just the filling fraction in the long-time limit. (c)-(d) Variation of the entanglement entropy $S_A$ and number fluctuations in the subsystem $\delta^2 N_A$ respectively with time(in units of $J^{-1}$) for increasing $\sigma$. The thick solid line shows linear dependence on time for comparison. For all the log-log plots $N=2048$, $L=N/2$ and number of disorder realizations is $100$ for fermions with filling fraction $0.1$ .}
\label{many}
\end{figure}

We also study the expansion dynamics of a cloud of fermions of a given filling and initial state in which fermions sit around the center of the lattice. This type of initial state can be prepared by switching on a trap potential and suddenly switching it off to study the evolution of the system under the quenched Hamiltonian. We calculate the expansion of the width of the many-particle cloud, which can be quantified by\cite{schneider2012}
\begin{equation}
w(t) = \sqrt{\frac{1}{N_p} \sum\limits_{i} (i-i_0)^2 \langle n_i(t)\rangle - \frac{1}{N_p} \sum\limits_{i} (i-i_0)^2 \langle n_i(0)\rangle},
\end{equation} 
where $N_p$ is the total number of particles and $\langle n_i\rangle$ is the average occupation at site $i$ whereas $i_0$ is the center of the lattice. Simultaneously another quantity $n_{loc}$, which is the sum of the occupation densities at the initially occupied sites, is also investigated as a function of time. This quantity is defined as\cite{masud}:
\begin{equation}
n_{loc}(t)= \frac{1}{N_p}\sum\limits_{i}^{in. occ.} \langle n_i(t)\rangle. 
\end{equation}
The width of the many-particle wavepacket $w$ in different phases is
shown in Fig.\ref{many}(a) and it shows the same qualitative feature
as $w^{sp}$. The variation of $n_{loc}$ with time nicely matches with
the dynamics of $w$ [Fig.\ref{many}(b)]. It decreases rapidly to the
saturation value, which is filling fraction in the delocalized phase
and barely changes in the localized phase. In the
  quasi-localized phase, it saturates to an intermediate value, which
  increases abruptly as $\sigma$ increases in the same phase. Also
we calculate the entanglement entropy for the same initially localized
many-particle state by choosing a subsystem of $L=\frac{N}{2}$
consecutive sites, whose center coincides with the center of the
lattice. It shows the same qualitative feature as
$S^{sp}_A$[compare Fig.~\ref{single}(a) and Fig.\ref{many}(c)]. Therefore, 
similar to the single-particle picture, there is a contrast between charge transport and entanglement
propagation in the many-particle picture. The number fluctuations also
show similar dependence on time but it is smoother than $S_A$
[Fig.\ref{many}(d)]. The roughness of $S_A$ and $S^{sp}_A$ may be an
artifact to the special choice of the subsystem. This whole analysis
has been carried out at a filling of $0.1$; however, we have verified
that there is no qualitative dependence of these results on the
filling fraction since there is no mobility edge in the energy
spectra.
\begin{figure}[h!]
\centering
\includegraphics[width=4.275 cm,height=4.0 cm]{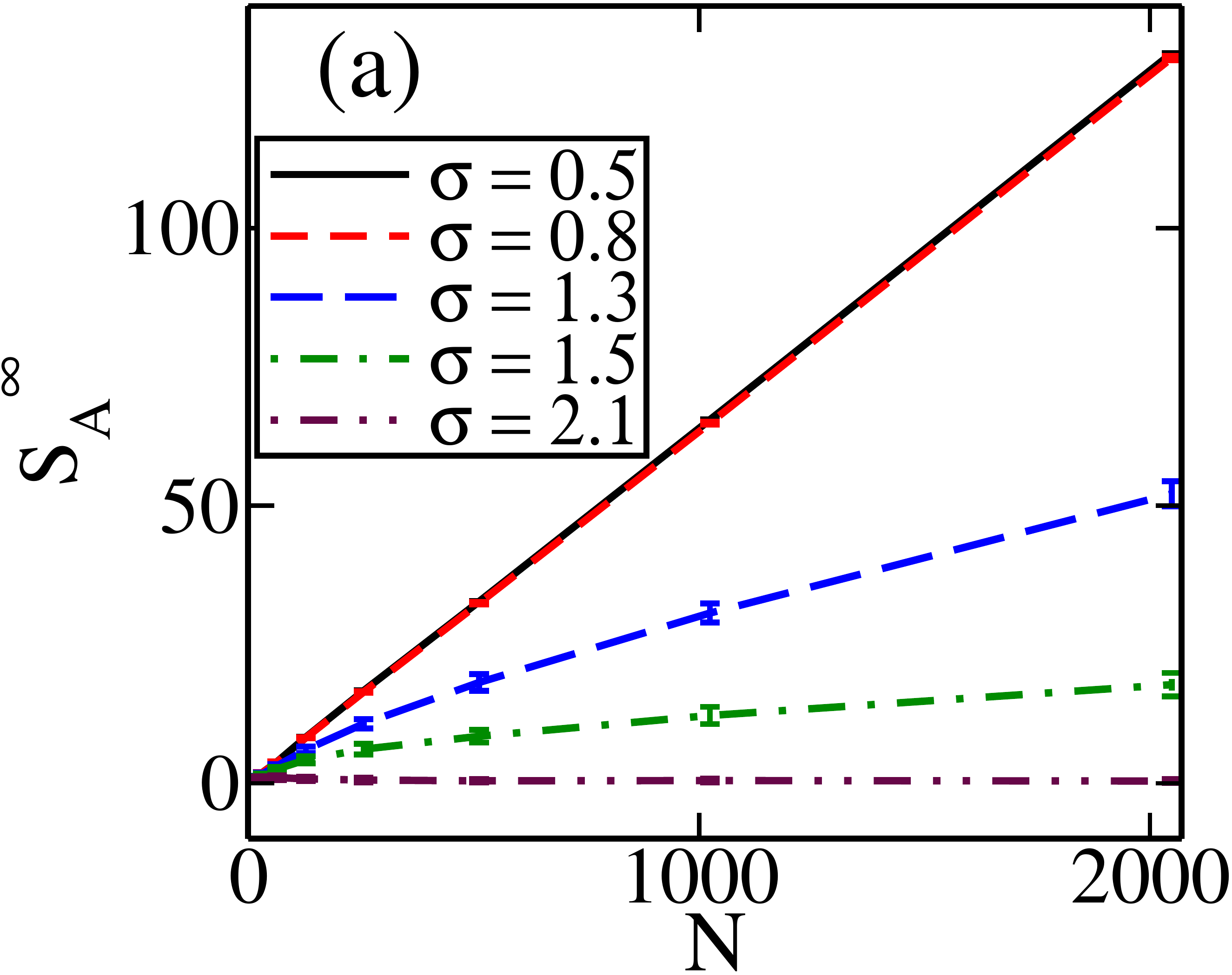}
\includegraphics[width=4.275 cm,height=4.0 cm]{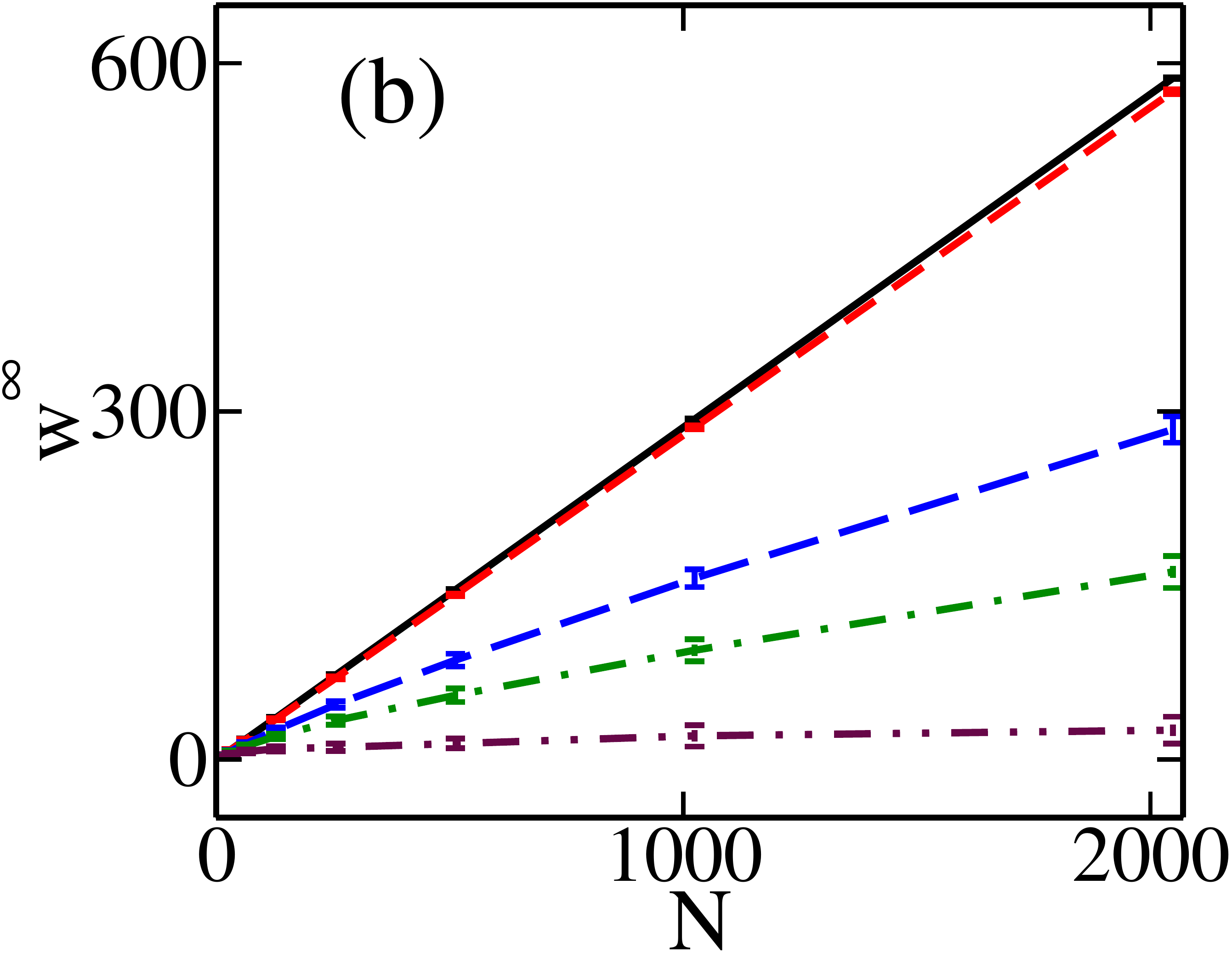}
\includegraphics[width=4.275 cm,height=4.0 cm]{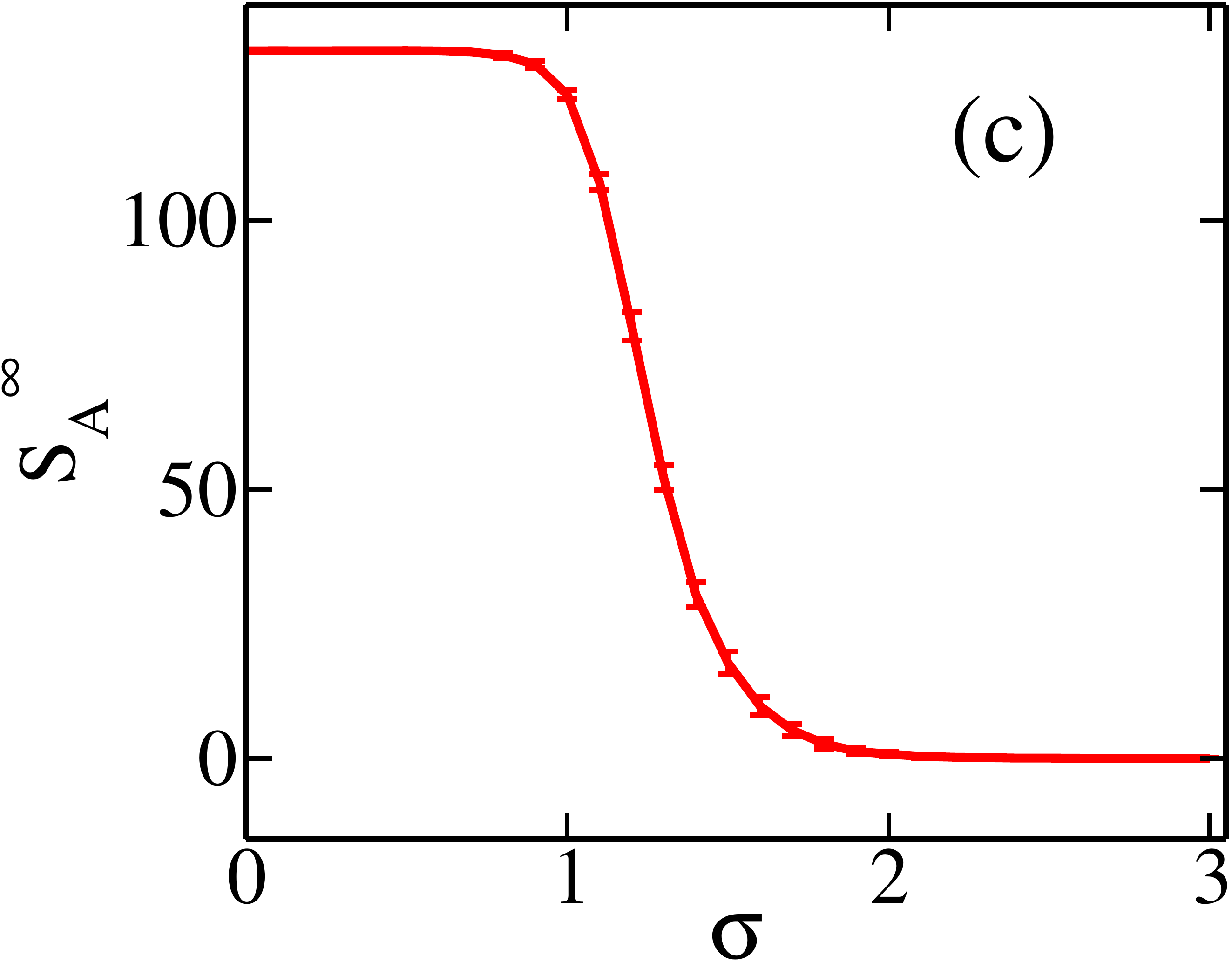}
\includegraphics[width=4.275 cm,height=4.0 cm]{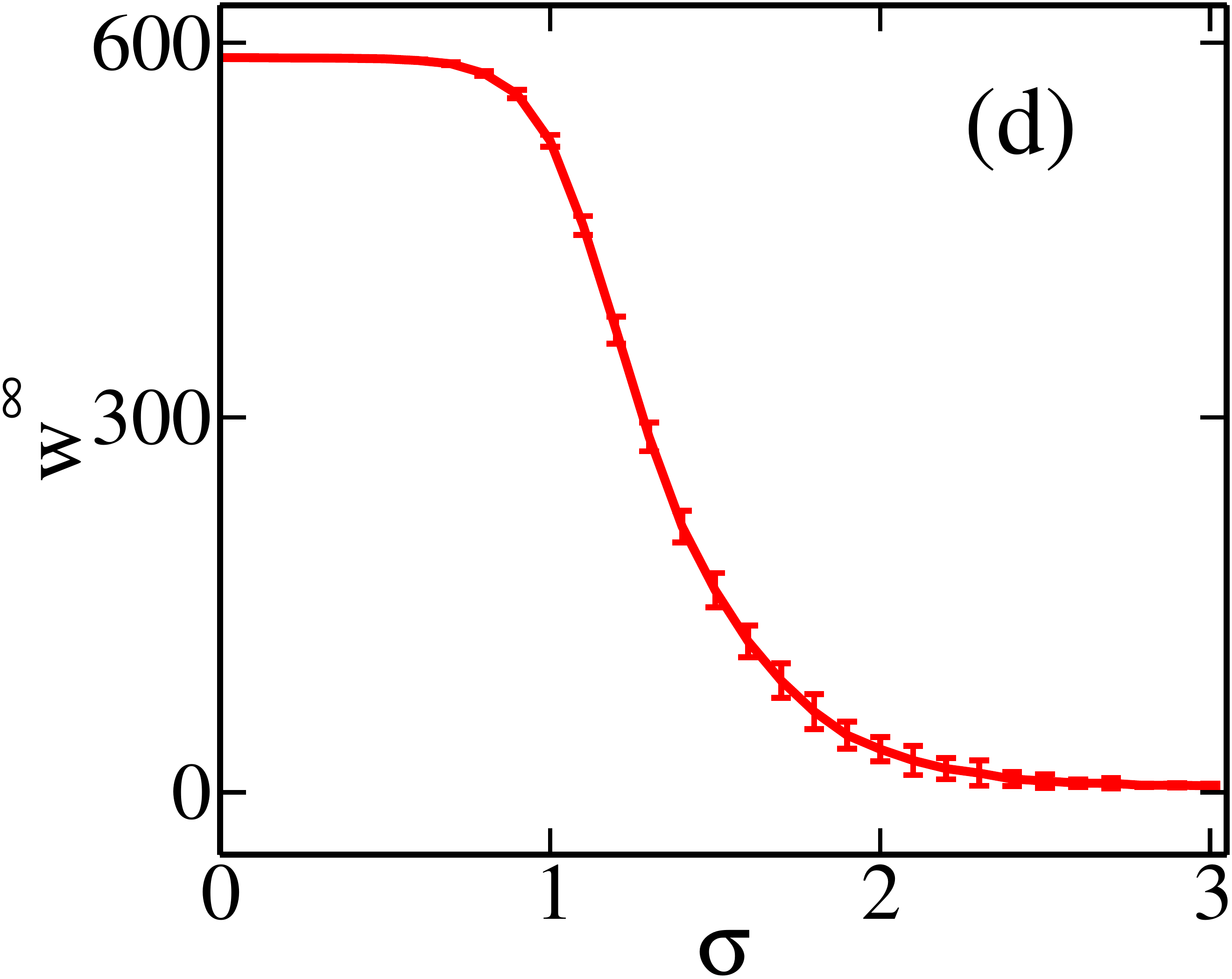}
\caption{(a)-(b) Scaling of the saturation values of the entanglement entropy $S_A^{\infty}$ and width of the many-particle wave-packet $w^{\infty}$ respectively with the system size $N$. (c)-(d) Variation of $S_A^{\infty}$ and $w^{\infty}$ respectively with $\sigma$. For Fig.(c),(d) $N=2048$. For all the plots $L=N/2$ and number of disorder realizations is $100$ for fermions with filling fraction $0.1$ .}
\label{inf}
\end{figure}
The saturation values of the many-particle entanglement entropy and
the width of the wavepacket show similar variation with the system
sizes $N$ [Fig.\ref{inf}(a-b)]. In the delocalized phase both the
quantities go linearly with $N$ whereas in the quasi-localized phase
the dependence is sub-linear and they become almost independent of $N$
in the localized phase. This is quite expected as it
  reflects the sensitivity of the three phases to the boundaries of
  the system. The variation of these two quantities with $\sigma$ is
  shown in [Fig.\ref{inf}(c-d)] and they show similar dependences. In
  the delocalized phase both the quantities have almost constant and
  very high values whereas in the quasi-localized phase their values
  decrease abruptly with $\sigma$ and for large $\sigma$ in the localized phase, become tiny  and
  almost $\sigma$-independent.

\section{Conclusion}      
To summarize, in this paper we study many static and dynamical
quantities to investigate the link between the
delocalization-localization transition and entanglement of spinless
fermions in a random long-range hopping model. Within the system sizes
used for numerical analysis, the system shows a delocalized phase for
$\sigma<1$ and a localized phase for $\sigma>2$. One also obtains a
quasi-localized phase for $1<\sigma<2$, as reflected by the
level-spacing ratio and wave-packet dynamics, but this phase may
vanish in the thermodynamic limit as hinted in the plots of
level-spacing ratio for different system sizes. Scaling of the
entanglement entropy with subsystem size reveals strong area-law
violation in the delocalized phase whereas the (quasi)localized phase
seems to adhere (for larger subsystems) strictly to the area law. In addition 
to the eigenvalues of the entanglement Hamiltonian, the maximally
entangled mode or the zero mode of the entanglement Hamiltonian,
also captures the localization transition, despite it being a non-topological
system. The entanglement contour, which is constructed out of both the
eigenvalues and the eigenfunctions of the entanglement Hamiltonian,
gives a picture of the spatial distribution of entanglement inside the
subsystem and nicely explains the violation of the area-law in the
system. Particle-number fluctuations in the subsystem have similar
dependence on space and time as the entanglement entropy. The ratio of
these two quantities shows a sharp signature at the point of the
localization transition. However, the nature of this signature is dependent
on the model in question as it is different in the AAH model from our
long-range model. The distribution of the ratio of the entanglement
contour to the fluctuation contour is sharply peaked in the
delocalized phase but the peak starts vanishing as one goes into the
(quasi)localized phase.

Also we study quench dynamics and wave-packet dynamics of fermions at
the single-particle and many-particle levels. At both the levels the
entanglement propagation and the charge transport show a sharp
contrast. Entanglement entropy shows super-ballistic behavior both in the
delocalized phase and the (quasi)localized phase, although this appears
only as a transient in the latter. This super-ballistic behavior is
attributed to the picture based on the Lieb-Robinson bounds for the spreading of correlation post a global quench. Contrastingly, the width of the wave-packet varies ballistically with time in the delocalized phase
while in the (quasi)localized phase after ballistic transient it
shows a sub-ballistic behavior with time before it saturates. In a
short-range model with disorder, the light cone picture is valid, and therefore 
the time dependence of entanglement entropy is always sub-ballistic in general. However, in our model
long-range couplings give rise to super-ballistic behavior. 
The saturation values of the width and entanglement entropy
show similar dependence as a function of the system size and $\sigma$
reflecting the presence of three phases in finite systems.

In our study, we have been able to explain the strong area law
violation in our long-range model by implementing the idea of
entanglement contour and connect them to the
delocalization-localization transition in the system by studying
quench and wave-packet dynamics. We hope that our results regarding
the relationship between entanglement entropy and number fluctuations
will help boost the possibility of indirect measurement of
entanglement in experiments. Also we have shown explicitly the
contrast between charge and entanglement transport, which is one of
the current topics of interest. As a future possibility, one can also
look for many-body localized phases in an interacting version of this
model. We hope that our work can trigger experimental studies of the disordered long-range model
in ongoing ionic trap experiments.

\section*{Acknowledgements} 
We are grateful to the High Performance Computing (HPC) facility at IISER Bhopal, where large-scale calculations in this project were run.
A.S is grateful to Simone Paganelli and Andrea Trombettoni for helpful discussions, and to SERB for the startup grant (File Number: YSS/2015/001696).
N.R acknowledges Sourin Das for bringing to his attention useful references, and CSIR-UGC, India for his Ph.D fellowship. 
\bibliography{refs}

\newpage

\begin{widetext}
\appendix

\begin{center}
{\bf Appendix}
\end{center}
Here we provide a detailed discussion about the methodologies used in the paper to calculate, namely, the single particle entanglement entropy, the fermionic entanglement entropy and non-equilibrium dynamics of the entanglement entropy.   
\section{Single particle Entanglement Entropy}\label{appA}
A normalized single particle eigenstate $\ket{\psi}$ can be expressed as,
\begin{equation}
\ket{\psi}=\sum\limits_{i\in A} \psi_i {c_i}^\dagger \ket{0}_A\otimes\ket{0}_B + \sum\limits_{i\in B} \psi_i\ket{0}_A\otimes{c_i}^\dagger\ket{0}_B,
\label{psi} 
\end{equation}

where $\ket{0}_{A}=\underset{i\in {A}}{\otimes}\Ket{0}_i$ and $\ket{0}_{B}=\underset{i\in {B}}{\otimes}\Ket{0}_i$.\\

We define $\ket{1}_{A}=\frac{1}{\sqrt{p_{A}}}\sum\limits_{i\in {A}} \psi_i {c_i}^\dagger \ket{0}_{A}$ and $\ket{1}_{B}=\frac{1}{\sqrt{p_{B}}}\sum\limits_{i\in {B}} \psi_i {c_i}^\dagger \ket{0}_{B}$.\\

Here $p_A=\sum\limits_{i\in A}|\psi_i|^2$; $p_B=\sum\limits_{i\in B}|\psi_i|^2$ and $p_A + p_B=1$.\\

Notice that ${\langle0|0\rangle}_A={\langle0|0\rangle}_B=1$ and ${\langle1|1\rangle}_A={\langle1|1\rangle}_B=1$.\\

We can now write Eq.\ref*{psi} as,
\begin{equation}
\ket{\psi}=\sqrt{p_A}\ket{1}_A\otimes\ket{0}_B + \sqrt{p_B}\ket{0}_A\otimes\ket{1}_B, 
\end{equation}

The density matrix of the full system $\rho^{sp}=\ket{\psi}\bra{\psi}$.\\

The reduced density matrix of subsystem A $\rho^{sp}_A=Tr_B[{\rho^{sp}}]$, 
which is given by,
\begin{equation}
\rho^{sp}_A=p_A\ket{1}_A\bra{1}_A + p_B\ket{0}_A\bra{0}_A, 
\end{equation}

Single particle entanglement entropy $S^{sp}_A=-Tr[\rho^{sp}_A\ln(\rho^{sp}_A)]$, which can be written as,
\begin{equation}
S^{sp}_A=-{p_A}\ln{p_A} -{p_B}\ln{p_B}. 
\end{equation}

\section{Fermionic Entanglement Entropy}\label{appB}
In the following, we explain the methodology to calculate entanglement entropy of $N_p$ non-interacting spinless fermions in the ground state of a 1D lattice of $N$ sites under periodic boundary condition. The generic Hamiltonian is given by,
\begin{equation}
H_1 =  \sum\limits_{i,j=1}^{N} t_{ij} c_{i}^{\dagger}c_{j} + h.c.
\end{equation}
The diagonal form of the Hamiltonian is given by,
\begin{equation}
H_1 =  \sum\limits_{k=1}^{N} \epsilon_{k} b_{k}^{\dagger}b_{k},
\end{equation} 
where $b_k=\sum\limits_{j=1}^{N}\psi_j(k) c_j$.\\
We calculate the entanglement entropy for the fermionic ground state, which is defined as,
\begin{equation}
\ket{\Psi_0}=\prod_{k=1}^{N_p}b_k^\dagger\ket{0}
\end{equation}
Due to Slater determinant structure of $\ket{\Psi_0}$, all higher correlations can be obtained by two point correlation $C_{ij}=\langle c_i^\dagger c_j\rangle$\cite{peschel2003calculation,peschel2009,peschel2012special}. 
The density matrix of the full system $\rho=\ket{\Psi_0}\bra{\Psi_0}$ and the reduced density matrix of subsystem A $\rho_{A}=Tr_{B}(\rho)$. By definition a one particle function, in this case two-point correlation in the subsystem, can be written as,
\begin{equation}
C_{ij}=Tr[\rho_A c_i^\dagger c_j]
\label{correlation}
\end{equation}
However, this is possible according to Wick's theorem only when the reduced density matrix is the exponential of free fermionic operator\cite{peschel2003calculation},
\begin{equation}
\rho_{A}=\frac{e^{-H_{A}}}{Z},
\end{equation}
where $H_{A}=\sum\limits_{i,j=1}^{L} H_{ij}^A c_{i}^{\dagger}c_{j}$ is called the
entanglement Hamiltonian, and $Z$ is obtained to satisfy the condition $Tr[\rho_{A}] = 1$.\\
The entanglement Hamiltonian can be written in the diagonal form as,
\begin{equation}
H_A=\sum\limits_{k=1}^{L} h_k a_k^\dagger a_k,
\end{equation}
where $a_k=\sum_{j=1}^{L} \phi_j(k) c_j$. The reduced density matrix is then given by,
\begin{equation}
\rho_A=\frac{exp[-\sum\limits_{k=1}^{L}h_k a_k^\dagger a_k]}{\prod\limits_{k=1}^{L}[1 + exp(-h_k)].}
\label{rho_A}
\end{equation}
Using Eq.\ref{rho_A}, we can write Eq.\ref{correlation} as,
\begin{equation}
C_{ij}=\sum\limits_{k=1}^{L} \phi^\ast_i(k){\phi}_j(k) \frac{1}{e^{h_k}+1}.
\end{equation}
This shows the matrices $C$ and $H_A$ share the eigenstate $\ket{\phi_k}$ and their eigenvalues are related by,
\begin{equation}
\lambda_k=\frac{1}{e^{h_k}+1},
\label{relation}
\end{equation}
where $\lambda_k$'s are eigenvalues of matrix $C$ in the subsystem.\\
The entanglement entropy $S_A=-Tr[\rho_A \ln(\rho_A)]$, which can be simplified~\cite{sharma2015landauer} using Eq.\ref{rho_A} and Eq.\ref{relation} as,
\begin{equation}
S_A=-\sum\limits_{k=1}^{L} [\lambda_k \log \lambda_k + (1-\lambda_k) \log (1-\lambda_k)],
\end{equation} 

\section{Non-equilibrium dynamics of fermionic Entanglement Entropy}\label{appC}
In this section we discuss how to calculate dynamics of fermionic Entanglement entropy in our model, under the Hamiltonian,$H_1$ and an initial many-particle state $\ket{\Psi_{in}}$, which is not the many-particle ground state of $H_1$. The Hamiltonian is given by,
\begin{equation}
\mathcal{H} =  \sum\limits_{i \neq j}^{N} t_{ij} c_{i}^{\dagger}c_{j} + h.c.,
\end{equation}
The Hamiltonian can be written in the diagonal form, which is given by,
\begin{equation}
\mathcal{H} =  \sum\limits_{k=1}^{N} \epsilon_{k} b_{k}^{\dagger}b_{k},
\end{equation} 
where $b_k=\sum\limits_{j=1}^{N}\psi_j(k) c_j$. Assuming $\hbar=1$, the time evolution of the Heisenberg operators $b_k(t)$ is given by,
\begin{align}
\dot{b}_k&=\frac{1}{i}[b_k,\mathcal{H}] \nonumber\\
&=\frac{1}{i}\epsilon_k b_k,
\end{align}
Hence, $b_k(t)=e^{-i\epsilon_k t}b_k(0)$.\\ 
Here, for example, we consider a density-wave(DW) type of initial state, defined as,
\begin{equation}
\ket{\Psi_{in}}=c_2^\dagger c_4^\dagger....c_N^\dagger \ket{0},
\end{equation}
where lattice sites $N$ is even and number of fermions $N_p=N/2$.\\
In order to calculate the dynamics of the entanglement entropy one first constructs $L\times L$ correlation matrix within the subsystem A or B, i.e. $C_{ij}(t)=\bra{\Psi_{in}} c_i^\dagger(t) c_j(t) \ket{\Psi_{in}}$, where $i,j\in A$. Below we detail the the calculation of $\bra{\Psi_{in}}c_i^\dagger(t)c_j(t)\ket{\Psi_{in}}$.
\begin{align}
\bra{\Psi_{in}}c_i^\dagger(t)c_j(t)\ket{\Psi_{in}}&=\sum\limits_{k,k^\prime=1}^{N} \psi_i(k) \psi_j(k^\prime) e^{-i(\epsilon_k-\epsilon_{k^\prime})t} \bra{\Psi_{in}} b_{k\prime}^\dagger(0) b_k(0) \ket{\Psi_{in}} \nonumber\\
&=\sum\limits_{k,k^\prime=1}^{N} \sum\limits_{i^\prime,j^\prime=1}^{N} \psi_i(k) \psi_j(k^\prime) {\psi}^{\ast}_{i^\prime}(k) {\psi}^{\ast}_{j^\prime}(k^\prime) e^{-i(\epsilon_k-\epsilon_{k^\prime})t} \bra{\Psi_{in}} c_{i^\prime}^\dagger c_{j^\prime} \ket{\Psi_{in}} \nonumber\\
&=\sum\limits_{k,k^\prime=1}^{N} \sum\limits_{i^\prime=2,4,...}^{N} \psi_i(k) \psi_j(k^\prime) {\psi}^{\ast}_{i^\prime}(k) {\psi}^{\ast}_{i^\prime}(k^\prime) e^{-i(\epsilon_k-\epsilon_{k^\prime})t}.
\end{align}
The fermionic entanglement entropy $S_A(t)$ following the diagonalization of the $L\times L$ correlation matrix is given by,
\begin{equation}
S_A(t)=-\sum\limits_{m=1}^{L} [\lambda_m \log \lambda_m + (1-\lambda_m) \log (1-\lambda_m)],
\end{equation}
where $\lambda_m$'s are the eigenvalues of the subsystem correlation matrix. 
  
\end{widetext}

\end{document}